\documentclass[tran]{IEEEtran}
\usepackage{array}
\usepackage{supertabular}
\usepackage{indentfirst}
\usepackage{graphicx}
\usepackage{epsfig}
\usepackage{epstopdf}
\usepackage{amsmath}
\usepackage{amsfonts}
\usepackage{subfigure}
\usepackage{color,soul}
\usepackage{cite}
\usepackage{multirow}
\begin{document}
\sethlcolor{yellow}
\title{Node Flux-Linkage Synchronizing Control of Power Systems with 100\% Wind Power Generation Based on Capacitor Voltage Balancing Scheme}
\author{Yang Liu, Yanshan Chen, Yuexi Yang, Xiangyu Pei, Feng Ji
\thanks{Yang Liu, Yanshan Chen are with the School of Electric Power Engineering, South China University of Technology, Guangzhou, 510641, China. Xiangyu Pei is with the Huairou Laboratory, Beijing, China. Yuexi Yang and Feng Ji are with the State Grid Smart Grid Research Institute Co. Ltd., Beijing, 102209, China.}

}

\markboth{}%
{LIU \MakeLowercase{\textit{et al.}}: }

\maketitle

\begin{abstract}
This paper proposes a node flux-linkage synchronizing control method (NFSCM) for power systems with 100\% wind power generation based on a capacitor voltage balancing scheme (CVBS). Different from the conventional grid-forming controllers, NFSCM is designed to regulate inverters as virtual flux-linkage sources. Auto-synchronization of flux-linkage vectors is achieved through the CVBS-based NFSCM. The mismatch among the angular frequencies of flux-linkage vectors is eliminated by regulating the tracking errors of DC-link voltages, which establishes a negative feedback between the output frequency and active power of the inverter. NFSCM is adaptive to weak and strong grids. It avoids the excitation inrush currents in the step-up transformer of wind power generators. It also eliminates the DC components of the three-phase currents, and avoids low-frequency oscillations in active power. In order to limit the short-circuit current of inverters, a logic-based bang-bang funnel control (LBFC) is designed to control the switches of inverter bridges when over-current is detected. LBFC is able to restrict various fault currents within an acceptable range within the shortest time. LBFC and NFSCM are designed to operate in a switched manner according to a state-dependent switching strategy. Time-domain simulations were conducted on a 100\% wind power generation test system, and the performance of NFSCM and LBFC were investigated.
\end{abstract}


\section*{Nomenclature}\label{appen_nomenclature}
\footnotesize
\begin{table}[htbp]
\centering
\begin{tabular}{p{1cm}p{7cm}}
$\omega_{\mathrm{r}i}$ & Rotor speed of the SG of WPG$_{i}$ \\
$\omega_{\mathrm{r}\_\mathrm{ref}i}$ & Reference of $\omega_{\mathrm{r}i}$ \\
$K_{\mathrm{p}\_\mathrm{pitch}}$ & Proportional gain of the rotor speed control loop of wind turbine governors \\
$\beta_i$ & Pitch angle of the turbine blades of WPG$_{i}$ \\
$P_{\mathrm{me}i}$ & Power input to the capacitor of WPG$_{i}$ \\
$P_{\mathrm{me}\_\mathrm{ref}i}$ & Reference of $P_{\mathrm{me}i}$ \\
$P_{\mathrm{in}i}$ & Active power output of the SG of WPG$_{i}$ \\
$P_{\mathrm{in}\_\mathrm{ref}i}$ & The reference of $P_{\mathrm{in}i}$ \\
$K_{\mathrm{p}\_\mathrm{comp}}$ & Proportional gain of the proportional-integral (PI) controller of the power control loop of wind turbine governors \\
$K_{\mathrm{i}\_\mathrm{comp}}$ & Integral gain of the PI controller of the power control loop of wind turbine governors \\
$|\Psi_{i}|$ & Magnitude of the stator flux of the SG of WPG$_{i}$ \\
$|\Psi_{i}|_{\mathrm{ref}}$ & Reference of $|\Psi_{i}|$ \\
$E_{\mathrm{f}i}$ & Excitation voltage of the SG of WPG$_{i}$ \\
$K_{\mathrm{p}\_\mathrm{field}}$ & Proportional gain of the PI controller of the exciter of SGs \\
\end{tabular}
\end{table}

\footnotesize
\begin{table}[htbp]
\centering
\begin{tabular}{p{1cm}p{7cm}}
$K_{\mathrm{i}\_\mathrm{field}}$ & Integral gain of the PI controller of the exciter of SGs \\
$C$ & Capacity of the capacitor of WPGs \\
$V_{\mathrm{dc}i}$ & Capacitor voltage of the DC-link of WPG$_{i}$ \\
$V_{\mathrm{dc}\_\mathrm{nom}}$ & Nominal capacitor voltage of WPGs \\
$K_{\mathrm{pg}ji}$ & Proportional gains of the governor of WPG$_{i}$, $j=1,2,3$ \\
$I_{\mathrm{s}i}$ & Current injected by the energy storage of WPG$_{i}$ \\
$I_{\mathrm{s}\_\mathrm{ref}i}$ & Reference of $I_{\mathrm{s}i}$ \\
$L_{\mathrm{boost}}$ & Inductance of the boost converter connected outside the rectifier of the WPG\\
$D_{\mathrm{duty}i}$ & Duty cycle of the boost converter of WPG$_{i}$ \\
${v}_{ji}$ & voltage of phase $j$ of the PCCB of WPG$_{i}$, $j=\mathrm{a},\mathrm{b},\mathrm{c}$ \\
${i}_{\mathrm{L}ji}$ & phase $j$ inductor current of the $RL$ filter of WPG$_{i}$, $j=\mathrm{a},\mathrm{b},\mathrm{c}$ \\
${i}_{ji}$ & current of phase $j$ flowing out of the PCCB of WPG$_{i}$ , $j=\mathrm{a},\mathrm{b},\mathrm{c}$ \\
$f_{\mathrm{n}}$ & Nominal frequency of the system \\
$K_{\mathrm{I}i}$ & Gain of the phase angle control loop of the NFSCM of WPG$_{i}$ \\
$K_{\mathrm{e}i}$ & Gain of the exciter-mimicking loop of the NFSCM of WPG$_{i}$ \\
$V_{\mathrm{t}i}$ & Magnitude of the three-phase voltages measured on the PCCB of WPG$_{i}$ \\
$V_{\mathrm{t}\_\mathrm{ref}i}$ & Reference of $V_{\mathrm{t}i}$ \\
$L_{\mathrm{filter}}$ & Inductance of the $RL$ filter of WPGs \\
$R_{\mathrm{filter1}}$ & Resistance of the $RL$ filter of WPGs \\
$R_{\mathrm{filter2}}$ & Resistance of the $RC$ filter of WPGs \\
$C_{\mathrm{filter}}$ & Capacitance of the $RC$ filter of WPGs \\
$S_{jPi}$ & States of the upper bridge arm of the inverter of WPG$_{i}$\\
$S_{jNi}$ & States of the lower bridge arm of the inverter of WPG$_{i}$ \\
$S^{\mathrm{b}}_{ji}$ & Switching command for the upper bridge arm of phase $j$ generated by LBFC of WPG$_{i}$ \\
$S_{ji}$ & Switching command for the upper bridge arm of phase $j$ generated by NFSCM of WPG$_{i}$ \\
$P_{\mathrm{mn}i}$ & Nominal mechanical power of WPG$_{i}$ \\
$P_{\mathrm{n}i}$ & Nominal electrical power of WPG$_{i}$ \\
$V_{\mathrm{n}}$ & Nominal line-to-line voltage of the stator of the SG of WPGs \\
$V_{\mathrm{tn}}$ & Nominal line-to-line voltage of the PCCB of WPGs \\
$\mathbf{\Psi}_{\mathrm{t}i}$ & Flux-linkage vector of the PCCB of WPG$_{i}$ \\
$\mathbf{\Psi}^*_{\mathrm{t}i}$ & Reference of $\mathbf{\Psi}_{\mathrm{t}i}$ \\
$\theta^{{\Psi}*}_{i}$ & Phase angle reference of the flux-linkage vector of the PCCB of WPG$_{i}$ \\
$\Delta\theta^{{\Psi}*}_{i}$ & Output of the phase angle control loop of the NFSCM of WPG$_{i}$ \\
${\it{\Psi}}^{*}_{\mathrm{t}i}$ & Magnitude reference of the flux-linkage vector of the PCCB of WPG$_{i}$ \\
${\it{\Psi}}_{ji}$ & Reference of the flux-linkage of phase $j$ of the PCCB of WPG$_{i}$ \\
${\it{\Psi}}^*_{ji}$ & Reference of ${\it{\Psi}}_{ji}$ \\
$e_{ji}(t)$ & Tracking error of the fault current of phase $j$ of the inverter of WPG$_{i}$ \\
$\varphi^+_{ji}$ & Upper bound of $e_{ji}$ \\
$\varphi^-_{ji}$ & Lower bound of $e_{ji}$ \\
$q_{ji}(t)$ & Logic output of the LBFC of phase $j$ of WPG$_{i}$ \\
$\gamma_{i}$ & Magnitude threshold in the switching strategy of the controllers of WPG$_{i}$\\
$\tau_{i}$ & Time threshold in the switching strategy of the controllers of WPG$_{i}$\\
$\mathcal{T}$ & Output of the switching strategy of the controllers of WPGs\\[4pt]
$\bar{(\cdot)}$ & per unit value of a variable \\
\end{tabular}
\end{table}

\normalsize
\section{Introduction}
Wind power and solar generation have been growing at an unprecedented rate in the last decade. 100\% renewable supplied power grids are the ultimate prototype for power systems of mankind \cite{8721134}. Renewable power sources are connected to power grids through flexibly controlled power electronics inverters \cite{7505630}, which have completely different dynamics in comparison with synchronous generators (SGs) \cite{8742900}. How to establish stable frequency and voltages and how to guarantee the stability of such systems without any SG are open problems to be studied \cite{7497697}. Up to now, there is an abundance of work concerning the operation and control of power systems with high penetration rates of renewable power, most of which focuses on the control techniques of inverters.

Control methods for inverters can be categorized by grid-following and grid-forming depending on their synchronizing rules \cite{6200347}. Grid-following control schemes require a phase-locked loop (PLL) to measure the phase and frequency of the point of common coupling bus (PCCB) voltage. Their stable operation relies greatly on the stiffness of the external power grid, and they are suitable for the strong-grid condition \cite{9173757}. In contrast, most of the grid-forming control strategies do not need a PLL, and generate synchronized voltage and frequency references for inverters based on specific $P/f$ negative feedback mechanisms. Grid-forming inverters are suitable for weak grids, and tend to be unstable under strong-grid conditions. Virtual synchronous generator (VSG), virtual oscillator control (VOC), and capacitor voltage synchronizing control (CVSC) are typical grid-forming control methods available by now.

VSG emulates the natural response of SGs to synchronize with the external power grid, and offers virtual inertia and damping properties \cite{4596800}. VSG with an alternating moment of inertia was proposed in \cite{6919271}, which was expected to offer extra damping for oscillations of power systems. To address the potential oscillations introduced by multi-VSGs, \cite{9778180} suggested an adaptive virtual impedance scheme to realize the auto-adjustment of connection strength of inverters. Since VSG is a control-enabled SG, potential instability is foreseen when the control fails to coordinate with the prime power supplier and the external power source \cite{9663556}. Moreover, extra current control loops must be used if current regulation is desired in VSG, which significantly increases the risk of broadband resonance.

VOC employs a virtual oscillator to generate voltage reference signals for inverters, which allows inverters to accomplish self-synchronization with each other when operating in parallel \cite{7317584}. Nevertheless, neither the frequency of the reference signal nor that of the inverter can be adjusted automatically, resulting in the limited application of VOC \cite{8456529}. CVSC was first proposed for a 100MW-level power system in \cite{2018Capacitor}, in which the voltage of the DC-link was employed as the intermediate variable to establish the negative feedback of $P/f$. The method was further extended for the 100\% wind power generation system in \cite{liu2023capacitor}. Both of the CVSC above controlled inverters as voltage sources, i.e., three-phase voltage reference signals were produced to regulate the inverter. Extra current control loops are required to guarantee the current quality of the inverter, which would introduce broadband resonance problems as well.

Following the route of \cite{liu2023capacitor}, this paper proposes to regulate inverters as flux-linkage sources instead of voltage sources. Node flux-linkage is a more fundamental circuit variable compared with node voltage and branch current. The dynamics of node voltage and that of branch current can be unified with that of node flux-linkage, since node voltage is a (0,1) tensor of node flux-linkage and branch current is another way of expressing the drop of node flux-linkage on inductors. To realize the self-synchronizing of different flux-linkage inverters, a CVBS-based NFSCM is proposed to generate flux-linkage reference vectors for inverters via the voltage motion equations of DC-link capacitors. Voltage synchronization is achieved through flux-linkage synchronization, and the potential risk of inrush current is eliminated through flux-linkage control. Capacitors of wind power generators (WPGs) absorb or release energy in a coordinated manner with the load changes of the external power grid, such that inertial response is provided. Automatically load-sharing and the active power-capacitor voltage droop characteristic are fulfilled by a governor built within CVBS. Meanwhile, LBFC is designed to suppress the fault current of inverters, and it operates in a switched manner with NFSCM.

To summarise, this paper is structured as follows. Section II presents a description of the test power system of 100\% wind power generation
investigated in this paper, as well as the design of CVBS-based NFSCM. Section III derives LBFC for fault current rejection as well as the switching strategy for LBFC and NFSCM. Comparative simulation of the test system under the control of NFSCM combined with LBFC and AC voltage synchronizing control method (AVSCM) under multiple disturbance scenarios are given in Section IV. Based on the simulation results, conclusions are drawn in Section V.

\begin{figure}[!h]
\centering
\includegraphics[width=0.49\textwidth]{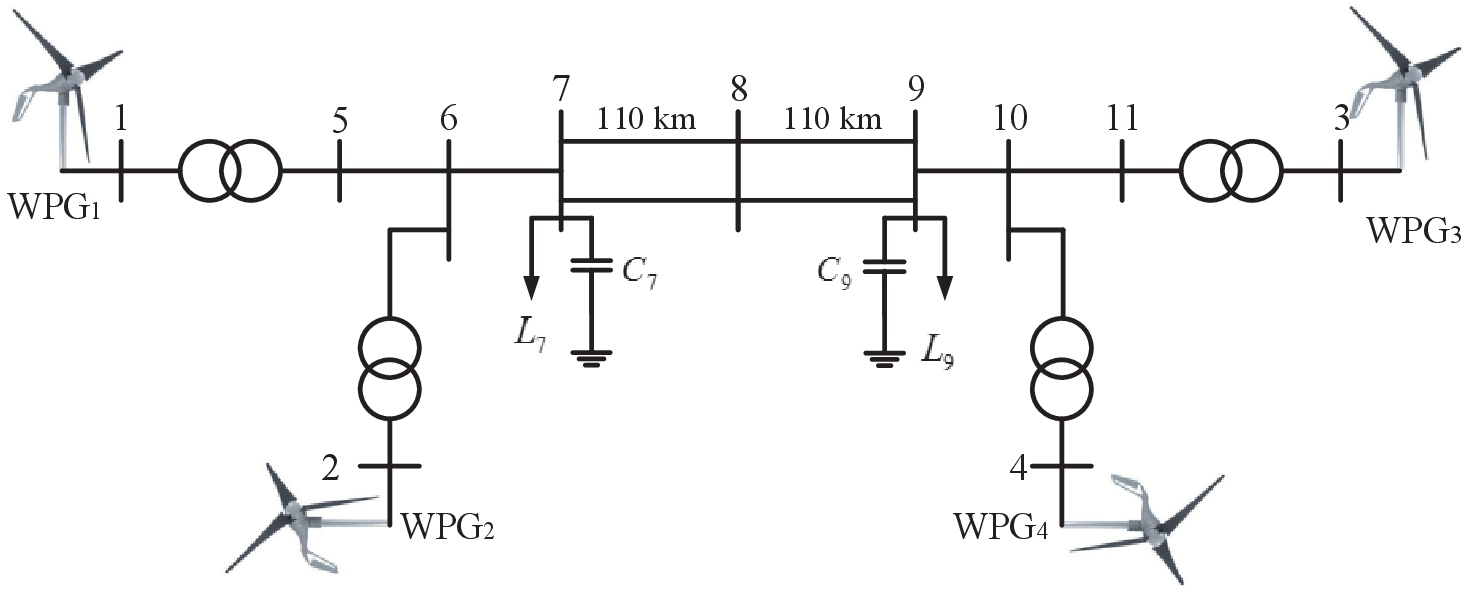}
\caption{Layout of a test power system with 100\% wind power generation.}
\label{fig_system_layout}
\end{figure}
\section{Designing CVBS-based NFSCM for A Test System With 100\% Wind Power Generation}\label{section_description_PEIPS_investigated}
\subsection{Description of The Test Power System With 100\% Wind Power Generation}\label{subsection_PEIPS_description}
Fig. \ref{fig_system_layout} illustrates the test power system with 100\% wind power generation studied in this paper, which is modified based on the Kundur four-machine two-area system \cite{kundur1994power}. As shown in Fig. \ref{fig_system_control}, the four SGs of the original system are swapped out for four 889 MW wind farms, each of which is simulated using an integrated model of a full-scale WPG with a 300 MW energy storage. Via DC-DC converters, energy storages are linked to the WPG capacitors. The specifications for transformers, loads, and transmission lines match those in \cite{kundur1994power}. The four WPGs configurations share the same configuration.
\subsection{Design of CVBS-based NFSCM}
WPGs controlled by NFSCM are expected to offer threefold performance: (1) establish voltage and frequency, and synchronize with the external power grid automatically; (2) provide inertial response; (3) offer load sharing for power balance. The capacitor voltage of the DC-link is a state variable that reflects the dynamic balance state of the power generated by the SG and that soaked by the grid-side inverter. Hence, it is chosen as an intermediate variable to build the negative feedback of $P/f$ for self-synchronization.

Dynamics of the capacitor voltage of the DC-link of WPG$_{i}$ can be described by
\begin{equation}\label{equ_capacitor_voltage_motion_equation}
CV_{\mathrm{dc}i}\frac{\mathrm{d}V_{\mathrm{dc}i}}{\mathrm{d}t}=P_{\mathrm{me}i}-P_{\mathrm{e}i}
\end{equation}
where $P_{\mathrm{me}i}=P_{\mathrm{in}i}+I_{\mathrm{s}i}V_{\mathrm{dc}i}$. Since the internal voltage vector of the SG is directly coupled with its rotor speed, the rotational speed of the internal voltage vector follows the equations of motion of the rotor \cite{kundur1994power}. It can be seen from (\ref{equ_capacitor_voltage_motion_equation}) that the capacitor voltage of an inverter exhibits the same dynamics as the rotor speed of an SG. Although WPG$_{i}$ does not have a rotating mass, it can be enabled to resemble the swing dynamics of an SG in frequency-disturbed events, on condition that the rotational speed of its internal voltage vector is governed by (\ref{equ_capacitor_voltage_motion_equation}).
\begin{figure}[t]
\centering
\includegraphics[width=0.5\textwidth]{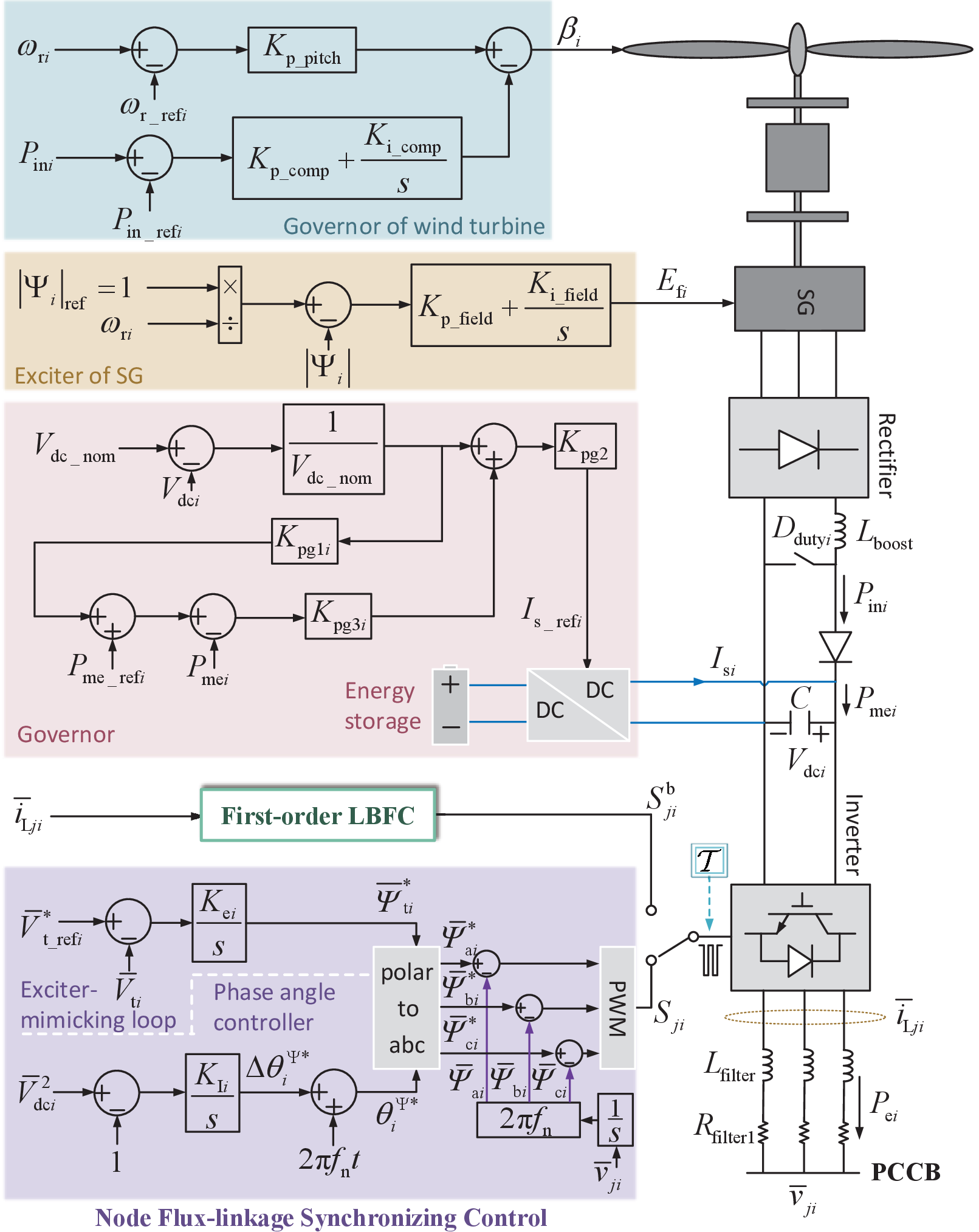}
\caption{The control system of WPG$_{i}$.}
\label{fig_system_control}
\end{figure}

The notion of flux-linkage is extended to circuit nodes in this paper. For example, the vector of the flux-linkage of WPG$_{i}$'s PCCB is written as $\mathbf{\Psi}_{\mathrm{t}i}(t)$, which is defined as $\mathbf{\Psi}_{\mathrm{t}i}(t)=\mathbf{\Psi}_{\mathrm{t}i}(t_{0})+\int^t_{t_0}\mathbf{V}_{\mathrm{t}i}(t)dt$. The obtained results meet the Faraday's law of electromagnetic induction, i.e., $\mathbf{V}_{\mathrm{t}i}(t)=p\mathbf{\Psi}_{\mathrm{t}i}(t)$, where $p$ is a differential operator, namely, $p=\frac{d}{dt}$. From this relationship, it is known that in steady state, $\mathbf{V}_{\mathrm{t}i}(t)$ is 90 degrees leading $\mathbf{\Psi}_{\mathrm{t}i}(t)$. Therefore, NFSCM is designed to synchronize the flux-linkage vector by regulating its rotational speed using negative feedback of the capacitor voltage of the DC-link.

Mimicking the perturbed form of the swing equation of an SG, the perturbed reference $\Delta\theta^{{\Psi}*}_{i}$ of the phase angle of the flux-linkage vector of the PCCB of WPG$_{i}$ is formulated via
\begin{equation}\label{equ_flux-linkage_small_signal}
\left\{
\begin{array}{l}
\displaystyle \frac{\mathrm{d}\Delta \theta^{{\Psi}*}_{i}}{\mathrm{d}t}=K_{\mathrm{I}i}\Delta \bar{V}^{2}_{\mathrm{dc}i}\\
\displaystyle \Delta\left(\bar{C}\bar{V}_{\mathrm{dc}i}\frac{\mathrm{d}\bar{V}_{\mathrm{dc}i}}{\mathrm{d}t}\right)=\Delta \bar{P}_{\mathrm{me}i}-\Delta \bar{P}_{\mathrm{e}i}
\end{array}
\right.
\end{equation}
where $\Delta\bar{V}^{2}_{\mathrm{dc}i}=(V_{\mathrm{dc}i}^{2}-V_{\mathrm{dc}\_\mathrm{nom}}^{2})/V_{\mathrm{dc}\_\mathrm{nom}}^{2}$, and $K_{\mathrm{I}i}$ is a gain designed for better transient performance. The phase angle reference of the flux-linkage vector of the PCCB in NFSCM is then generated by
\begin{equation}
\theta^{\Psi*}_{i} = \Delta\theta^{\Psi*}_{i} + \omega_{\mathrm{n}}t
\end{equation}
where $\omega_{\mathrm{n}}=2\pi f_{n}$.

An exciter-mimicking loop is designed to generate the magnitude reference $\bar{\it{\Psi}}^*_{\mathrm{t}i}$ for the flux-linkage vector with the voltage error of the PCCB of WPG$_{i}$ in NFSCM. The overall structure of NFSCM is as illustrated in Fig. \ref{fig_system_control}. The reference vector $\mathbf{\bar{\Psi}}^{*}_{\mathrm{t}i}$ of the flux-linkage is then denoted as
$\mathbf{\bar{\Psi}}^{*}_{\mathrm{t}i}={\it{\bar{\Psi}}}^{*}_{\mathrm{t}i}\angle\theta^{{\Psi}*}_{i}$. The three-phase references of the flux-linkages of the PCCB then can be obtained by polar to abc transformation, i.e.,
\begin{equation}
\begin{aligned}
& \bar{\it{\Psi}}^*_{\mathrm{a}i}=\bar{\it{\Psi}}^*_{\mathrm{t}i}\mathrm{cos}\theta^{\Psi*}_{i}\\
& \bar{\it{\Psi}}^*_{\mathrm{b}i}=\bar{\it{\Psi}}^*_{\mathrm{t}i}\mathrm{cos}(\theta^{\Psi*}_{i}-2\pi/3)\\
& \bar{\it{\Psi}}^*_{\mathrm{c}i}=\bar{\it{\Psi}}^*_{\mathrm{t}i}\mathrm{cos}(\theta^{\Psi*}_{i}+2\pi/3)\\
\end{aligned}
\end{equation}
The actual values of flux-linkage can be calculated by the integration of node voltages, i.e.,
\begin{equation}
\begin{aligned}
&\bar{\it{\Psi}}_{ji}(t)=\bar{\it{\Psi}}_{ji}(t_{0})+\int^{t}_{t_{0}}\bar{v}_{ji}(t)dt\\
\end{aligned}
\end{equation}
where $j\in\{\mathrm{a},\mathrm{b},\mathrm{c}\}$. The tracking errors of the three-phase flux-linkages are fed into the PWM module to generate the pulse control signals $S_{ji}$ for bridge arms, as shown in Fig. \ref{fig_system_control}.

Generation-load balance in the conventional SG-based power system is characterized by the synchronization of the rotor speed of SGs. In contrast, synchronization of the WPG-based power system controlled by NFSCM reveals that
\begin{equation}\label{equ_capacitor_synchronize}
\begin{aligned}
\frac{\mathrm{d}\Delta \theta^{{\Psi}*}_{i}}{\mathrm{d}t}-\frac{\mathrm{d}\Delta \theta^{{\Psi}*}_{j}}{\mathrm{d}t}=&K_{\mathrm{I}i}(\Delta \bar{V}^{2}_{\mathrm{dc}i}-\Delta \bar{V}^{2}_{\mathrm{dc}j})=0
\end{aligned}
\end{equation}
if $K_{\mathrm{I}i}$ of all WPGs takes the same value. Moreover, it has
\begin{equation}\label{equ_voltage_synchronization}
\begin{aligned}
&\Delta \bar{V}_{\mathrm{dc}i}-\Delta \bar{V}_{\mathrm{dc}j} = 0\\
&\bar{V}_{\mathrm{dc}i}-\bar{V}_{\mathrm{dc}j} = 0
\end{aligned}
\end{equation}
if the initial values of $\bar{V}_{\mathrm{dc}i}$ and $\bar{V}_{\mathrm{dc}j}$ are the same. Consequently, generation-load balance in the power system with 100\% WPG is characterized by the synchronization of the capacitor voltages of WPGs on the condition that $K_{\mathrm{I}i}\neq 0$. Combining (\ref{equ_flux-linkage_small_signal}) and (\ref{equ_capacitor_synchronize}), it can be found that the internal flux-linkage vector of the WPG controlled by CVBS-based NFSCM moves in the same manner as the internal voltage vector of an SG.


Analogous to SG-based conventional power systems, primary frequency regulation is needed in the 100\% wind power generation system for proper load sharing among WPGs. However, the primary frequency regulation in NFSCM is realized by regulating the post-disturbance error in the capacitor voltage. A governor is implemented as presented in Fig. \ref{fig_system_control}, which controls the power flowing through the capacitor using a parallel energy storage. Active power-capacitor voltage droop characteristic is realized by the $K_{\mathrm{pg}1}$ loop in the governor.

Compared with conventional power systems, in which constant frequency errors exist after primary frequency regulations, there would be constant capacitor voltage errors after the primary regulation of WPGs. Hence, additional power flow management is needed for long-term stability control of the 100\% wind power generation system, which is out of the scope of this paper. With respect to the generator and turbine controls of the WPG, a constant flux controller is applied for the excitation control of the SG of a WPG, and a wind turbine governor is implemented for mid-term rotor speed and active power management of SGs, as displayed in Fig. \ref{fig_system_control}. The configurations of these controllers are the same as in \cite{liu2023capacitor}.

\section{Design of Fault Current Rejection Control}\label{switching control}
\subsection{Logic-Based Bang\mbox{-}Bang Funnel Control (LBFC) Design}\label{LBFC control}

\begin{figure}[ht]
\centering
\includegraphics[width=0.4\textwidth]{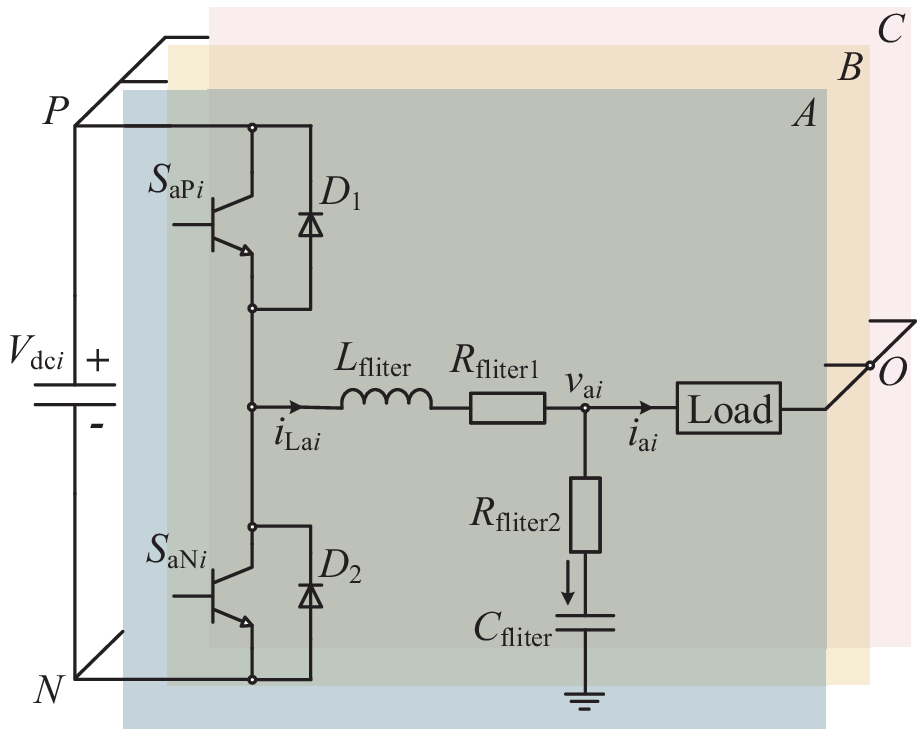}
\caption{Topology of the three-phase full-bridge inverter.}
\label{fig_topology_inverter}
\end{figure}

Topology of the three-phase full-bridge inverter of a full-scale WPG is as shown in Fig. \ref{fig_topology_inverter}. Switch $S_{jPi}$ and $S_{jNi}$ operate in a compensated manner, and $j\in\{\mathrm{a},\mathrm{b},\mathrm{c}\}$. Let $S_{jPi}$ be the state of the switch of the upper arm of phase $j$ and rewrite it as $S_{ji}^{\mathrm{b}}$. The domian of $S_{ji}^{\mathrm{b}}$ is $\{0,1\}$, and $S^{\mathrm{b}}_{ji}=1$ represents the on-state of the switch. Then the three-phase inverter can be modelled as follows.
\begin{equation}\label{equ_inverter modelling}
\left\{
\begin{aligned}
&L_{\mathrm{fliter}}\frac{di_{\mathrm{L}ji}}{dt}=V_{\mathrm{dc}i}S_{ji}^{\mathrm{b}}+u_{\mathrm{ON}i}-R_{\mathrm{fliter}}i_{\mathrm{L}ji}-v_{ji}\\
&C_{\mathrm{fliter}}\frac{du_{\mathrm{C}ji}}{dt}=i_{\mathrm{L}ji}-i_{ji}\\
&u_{\mathrm{ON}i}=-\frac{V_{\mathrm{dc}i}}{3}(S_{\mathrm{a}i}^{\mathrm{b}}+S_{\mathrm{b}i}^{\mathrm{b}}+S_{\mathrm{c}i}^{\mathrm{b}})
\end{aligned}
\right.
\end{equation}
where $u_{\mathrm{ON}i}$ donotes the voltage between point $O$ and $N$, $u_{\mathrm{C}ji}$ represents the voltage of the filter capacitor.

A first-order LBFC is employed for the fault current rejection of each phase, respectively. The LBFC was first proposed in our previous work \cite{8703197,8993693}. Given the tracking error of the fault current of phase $j$ as $e_{ji}(t)=i_{\mathrm{L}ji}-i^*_{\mathrm{L}ji}$, the switching logic of the first-order LBFC is
\begin{equation}\label{equ switching logic of LBFC}
\begin{aligned}
q_{ji}(t)=&\mathcal{G}(e_{ji}(t),\varphi^{+}_{ji},\varphi^{-}_{ji},q_{ji}(t-))\\
=&\mathcal{G}(e_{ji}(t),\varphi^{+}_{ji},\varphi^{-}_{ji},q_{ji}(t-))\\
:=&[e_{ji} \geq \varphi^{+}_{ji} \vee (e_{ji}>\varphi^{-}_{ji}\wedge q_{ji}(t-))]\\
q_{ji}(0-)&\in\{\mathrm{true},\mathrm{false}\}\\
\end{aligned}
\end{equation}
where $q_{ji}(t)\in\{\mathrm{true,false}\}$, $q_{ji}(t-):=\lim_{\varepsilon\rightarrow0}q_{ji}(t-\varepsilon)$. $q_{ji}(t)$ is the output of the switching logic (\ref{equ switching logic of LBFC}), and $i^*_{\mathrm{L}ji}=0$ is applied. LBFC is designed to drive $e_{ji}(t)$ into the error funnel defined by $[\varphi^{-}_{ji},\varphi^{+}_{ji}]$.
The control law of the first-order LBFC is given by
\begin{equation}
S_{ji}^{\mathrm{b}}(t)=\left\{\begin{aligned}&0, \quad \mathrm{if}\,q_{ji}(t)=\mathrm{true}\\
&1, \quad \mathrm{if}\,q_{ji}(t)=\mathrm{false}\end{aligned}\right.
\end{equation}

From (\ref{equ switching logic of LBFC}), it can be observed that the LBFC of each phase is dedicated to fault current control to prevent the inverter from over-current tripping. However, during normal operation, LBFCs cannot guarantee the synchronization of the inverter with the external power grid. Therefore, the three LBFCs should operate in a switched manner with NFSCM. The following switching strategy is designed for the proper switching between the two controllers.

\begin{figure}[h]
\centering
\includegraphics[width=0.49\textwidth]{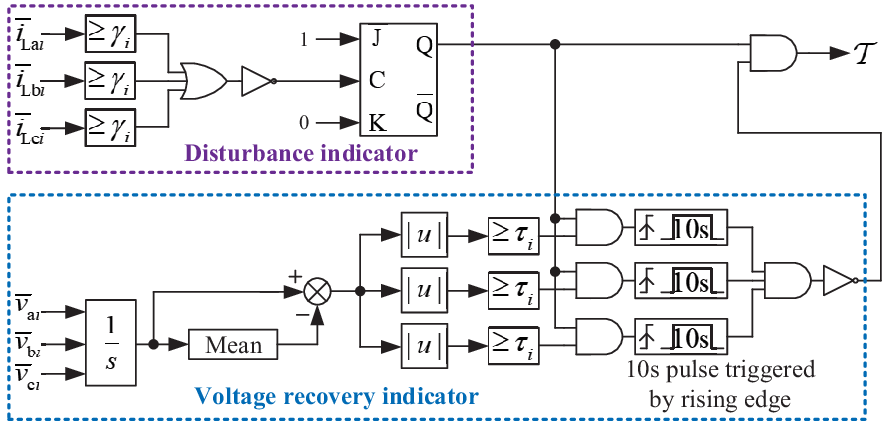}
\caption{Schematic of the switching strategy between LBFC and NFSCM.}
\label{fig_switchingrule}
\end{figure}
\subsection{Switching Strategy Design}\label{switching strategy}
A state-dependent switching strategy, depicted in Fig. \ref{fig_switchingrule}, is designed here. The strategy is composed of a disturbance indicator and a voltage recovery indicator. The disturbance indicator is dedicated to detect the over-current of the inverter, which is realized by checking $\bar{i}_{\mathrm{L}ji}\geq \gamma_{i}$. If this condition is satisfied, then the switching strategy outputs $\mathcal{T}=1$ and LBFCs are switched on operation. The voltage recovery indicator checks the absolute value of the difference between the integration of a phase voltage and its mean value. If a threshold $\tau_{i}$ is reached then it indicates that the voltage of PCCB has recovered. The switching strategy outputs $\mathcal{T}=0$ and NFSCM is switched on operation.

\section{Simulation Validation of the Test System}\label{section_time_domain_simulation_PEIPS}
In order to verify the effectiveness of the proposed NFSCM as well as LBFC in the stability control of 100\% wind power generation system, electromagnetic transient simulations were carried out with the test system shown in Fig. \ref{fig_system_layout}. In the test system, all WPGs are implemented with CVBS-based NFSCM. WPG$_{1}$ is also configured with LBFC for fault current rejection, and WPG$_{3}$ is chosen as the slake machine which does not have a governor installed. Parameters of WPGs and their control systems are given in Table \ref{tab_parameter}. Two typical operation scenarios were tested and the obtained results are as follows.

\begin{table}[htbp]
\centering
\caption{Parameters of the full-scale WPGs and their control systems.}
\label{tab_parameter}
\begin{tabular}{p{1cm}p{1.2cm}p{1cm}p{1.2cm}p{1cm}p{1cm}}
\hline
Parameter & Value & Parameter & Value & Parameter & Value \\
\hline
$P_{\mathrm{mn}i}$ & 800 MW & $P_{\mathrm{n}}$ & 889 MW & $V_{\mathrm{n}}$ & 730 V \\
$f_{\mathrm{n}}$ & 60 Hz & $V_{\mathrm{tn}}$ & 575 V & $L_{\mathrm{filter}}$ & 0.15 p.u. \\
$R_{\mathrm{filter}}$ & 0.003 p.u. & $V_{\mathrm{dc}\_\mathrm{nom}}$ & 1110 V & $C$ & 540 F \\
$L_{\mathrm{boost}}$ & 0.0012 H & $K_{\mathrm{p}\_\mathrm{pitch}}$ & 15 & $K_{\mathrm{p}\_\mathrm{comp}}$ & 1.5 \\
$K_{\mathrm{i}\_\mathrm{comp}}$ & 6 & $K_{\mathrm{p}\_\mathrm{field}}$ & 10 & $K_{\mathrm{i}\_\mathrm{field}}$ & 20 \\
$K_{\mathrm{pg1}}$ & 30 & $K_{\mathrm{pg2}}$ & 15 & $K_{\mathrm{pg3}}$ & 0.1 \\
$D_{\mathrm{duty}}$ & 0.19 & $K_{\mathrm{I}i}$ & 10 & $K_{\mathrm{e}i}$ & 0.2 \\
$\varphi^+_{ji}$ & 0.3 & $\varphi^-_{ji}$ & -0.3 & $\gamma_{i}$ & 1.75p.u. \\
$\tau_{i}$ & 0.5 & & & & \\
\hline
\end{tabular}
\end{table}

\begin{figure}[!h]
\centering
\includegraphics[width=0.485\textwidth]{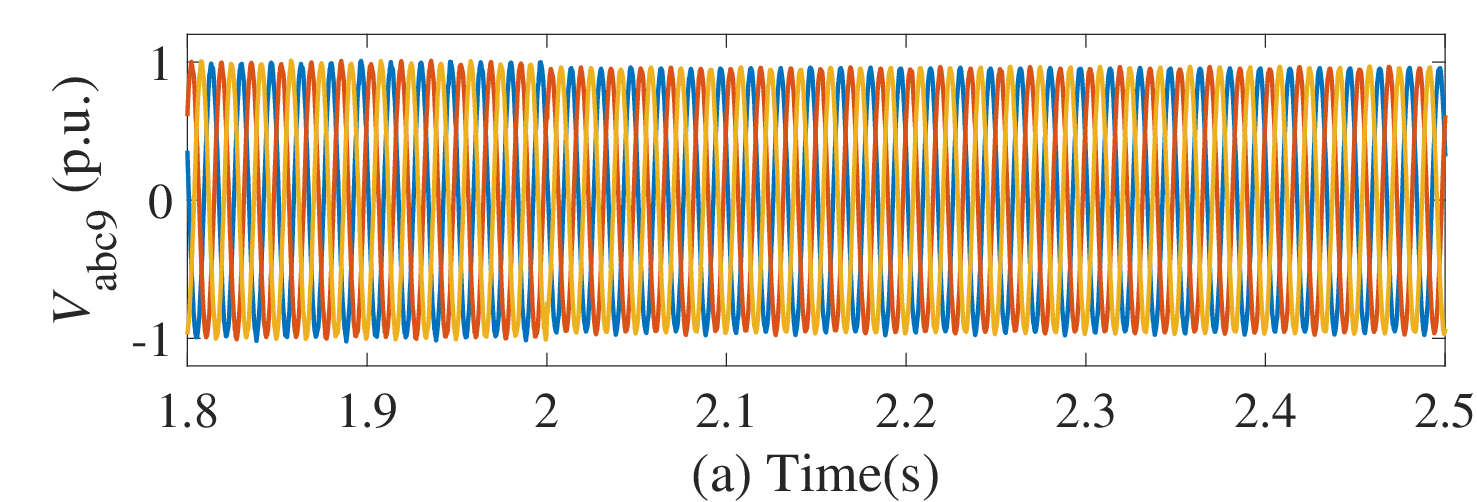}\\
\includegraphics[width=0.485\textwidth]{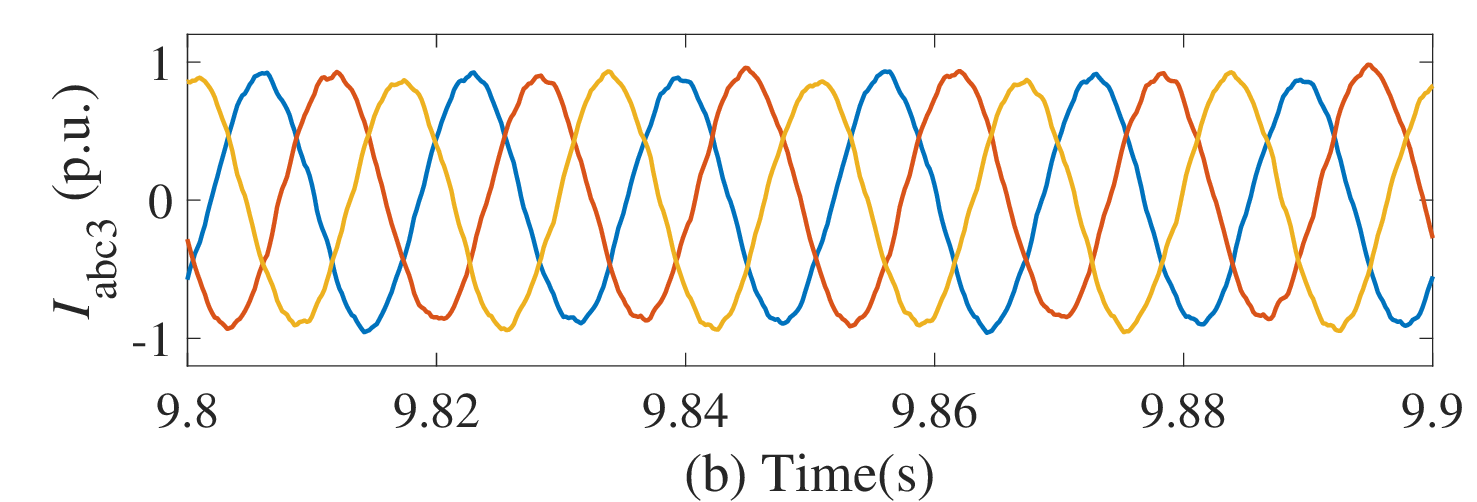}\\
\includegraphics[width=0.485\textwidth]{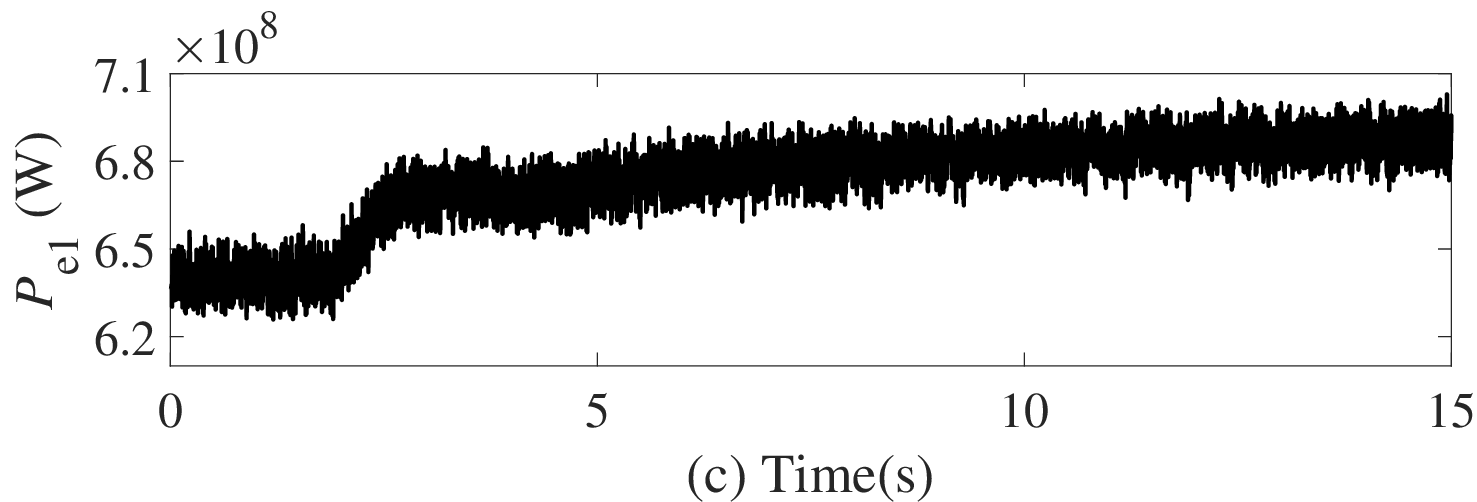}\\
\includegraphics[width=0.485\textwidth]{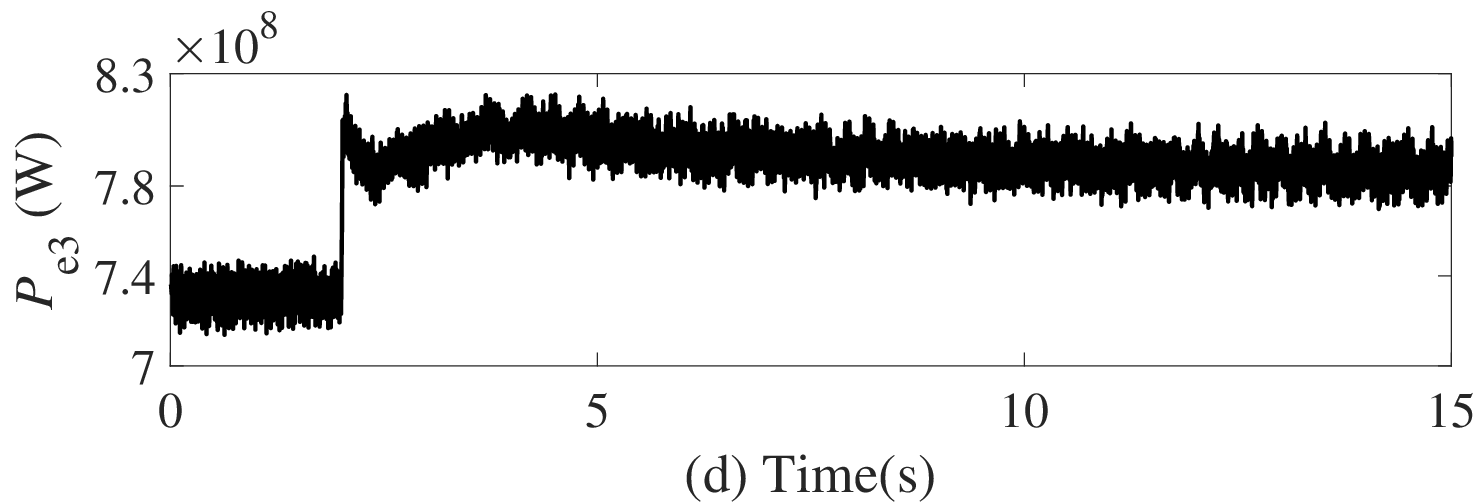}\\
\includegraphics[width=0.485\textwidth]{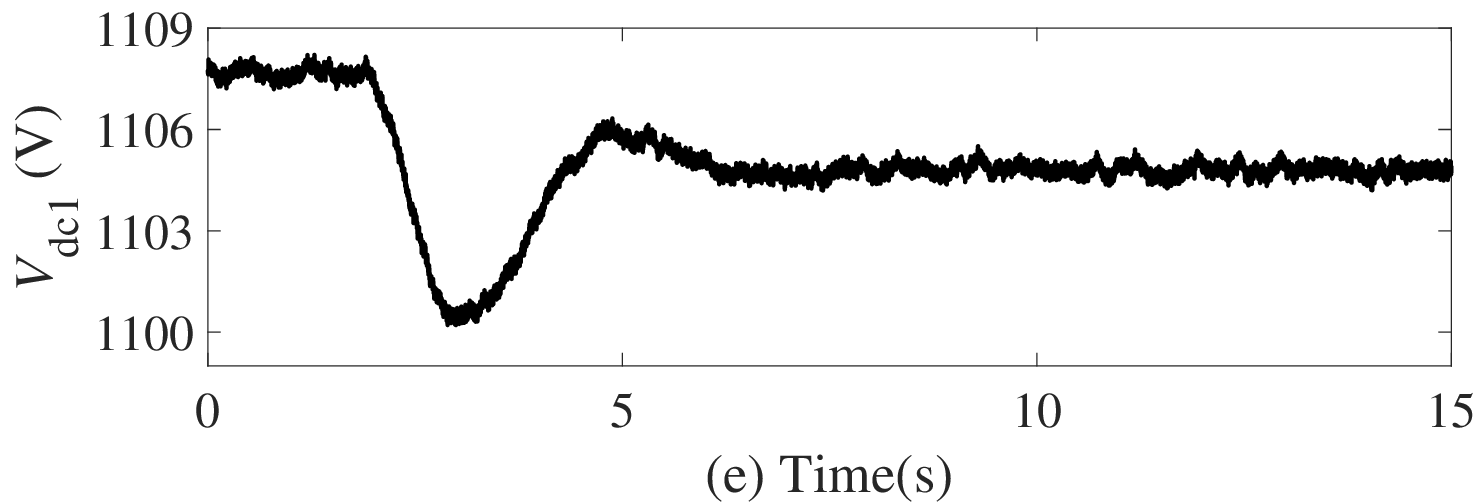}\\
\includegraphics[width=0.485\textwidth]{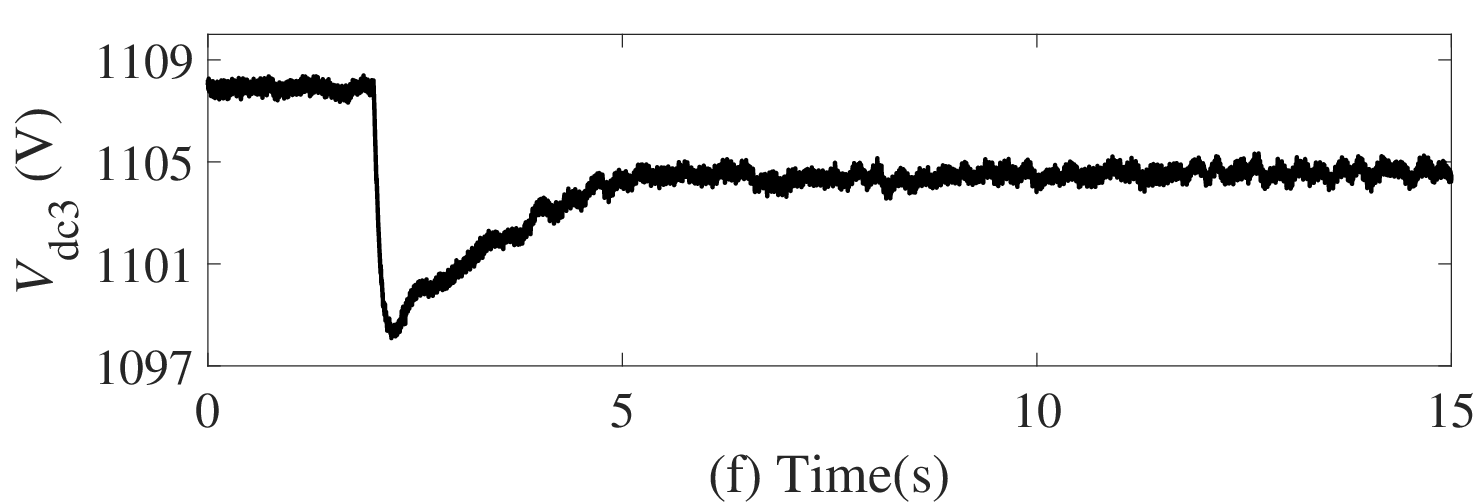}\\
\caption{Dynamics of WPG$_{1}$ and WPG$_{3}$ obtained in the case where a 400MW load was connected on load bus 9 at $t=2$s under the control of NFSCM ((a) Three-phase voltages measured on load bus 9 (b) Three-phase currents measured on generator bus 3 (c) Active power output of WPG$_{1}$ (d) Active power output of WPG$_{3}$ (e) Capacitor voltage of WPG$_{1}$ (f) Capacitor voltage of WPG$_{3}$).}
\label{fig_Load1}
\end{figure}
\subsection{A 400 MW Load Increase Occurred on Load Bus 9 of the Test System at $t=2$s}
The first case tested was a load-increase event. Specifically, a 400MW load was attached on load bus 9 at $t=2$s. Fig. \ref{fig_Load1} displays the dynamics of WPG$_{1}$ , WPG$_{3}$, and voltages of load bus 9. Due to the load increase, the magnitudes of load bus voltages dropped, and the three-phase voltages measured on load bus 9 were as shown in Fig. \ref{fig_Load1}(a). Consequently, the load current increased, which resulted in more active power output of WPG$_{1}$ and WPG$_{3}$ as illustrated in Fig. \ref{fig_Load1}(c) and Fig. \ref{fig_Load1}(d), respectively. From the results, it can be found that WPG$_{3}$ offered inertial response and primary frequency regulation, while WPG$_{1}$ only took part in the primary frequency regulation. The role that a WPG plays in the inertial response process is determined by the natural electric distance between the disturbed node and the WPG. Hence, WPG$_{3}$ offered step-up active power, while WPG$_{1}$ did not. This meets the phenomenon in SG-based power systems. The primary frequency regulation of WPGs was mainly achieved by the energy storage connected on the DC-link.

\begin{figure}[!t]
\centering
\includegraphics[width=0.485\textwidth]{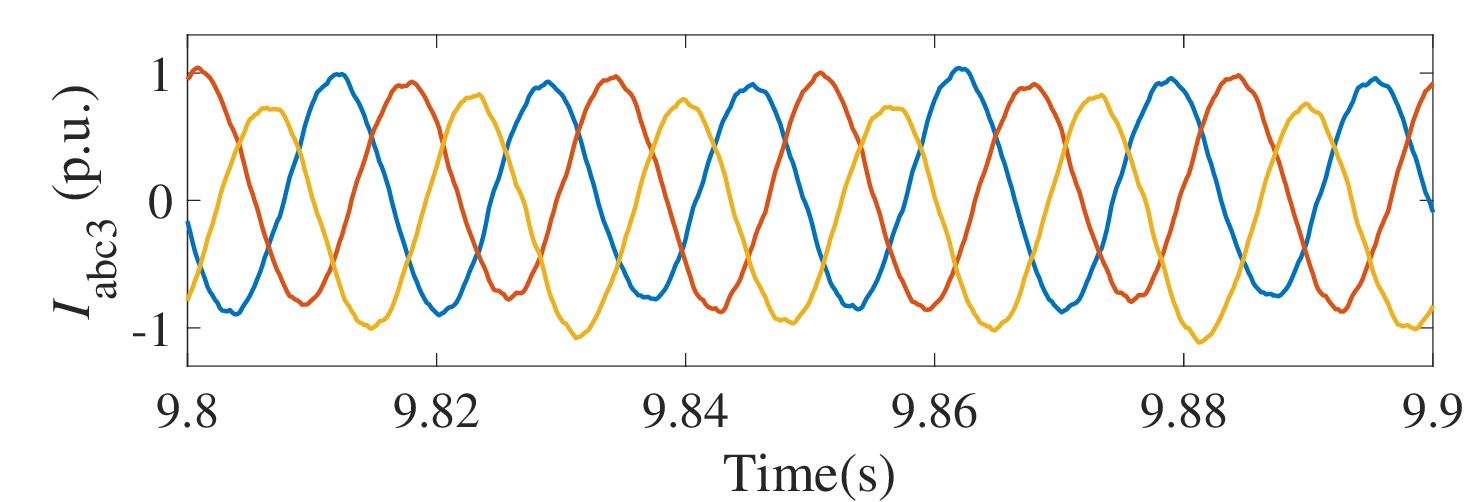}\\
\caption{Three-phase currents measured on generator bus 3 in the case where a 400MW load was connected on load bus 9 at $t=2$s under the control of AVSCM.}
\label{fig_Load2}
\end{figure}

As depicted in Fig. \ref{fig_Load1}(f), the capacitor voltage of WPG$_{3}$ dropped due to the upsurge of its active power. In comparison, the capacitor voltage of WPG$_{1}$ dropped moderately after the load increase occurred, as depicted in Fig. \ref{fig_Load1}(e). Although the capacitor voltages of WPGs did not drop to the same nadir, they would all converge to a new common operating level under the control of NFSCM and achieve novel synchronization. Three-phase AC currents measured on generator bus 3 under the control of AVSCM, as detailed in Fig. \ref{fig_Load2}, had a significant DC component. By implementing NFSCM, this issue was solved, and the three-phase currents displayed in Fig. \ref{fig_Load1}(b) became symmetrical and DC component-free.

\subsection{Inrush Currents in No-load Operation and A Three-phase-to-ground Fault Occurred on the Test System}

\begin{figure}[!t]
\centering
\includegraphics[width=0.485\textwidth]{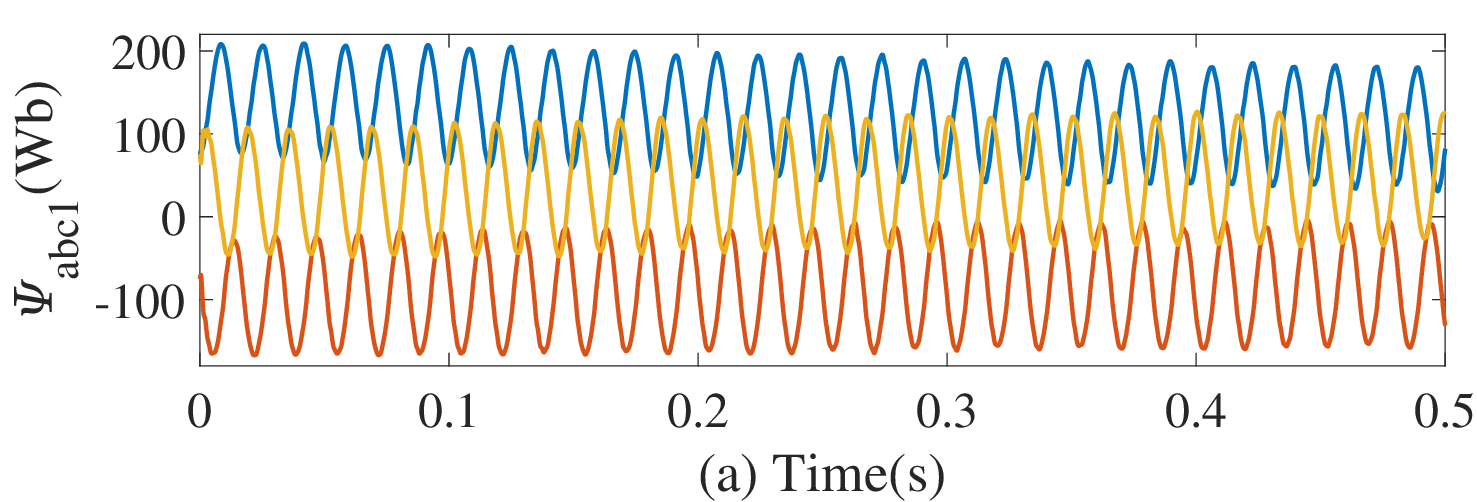}\\
\includegraphics[width=0.485\textwidth]{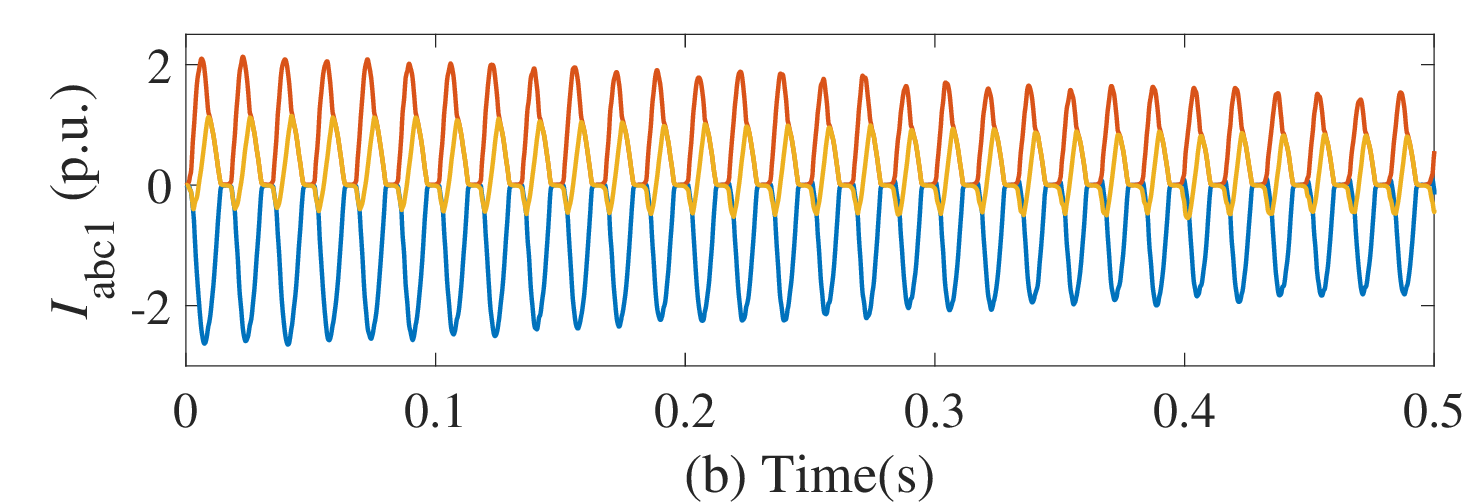}\\
\includegraphics[width=0.485\textwidth]{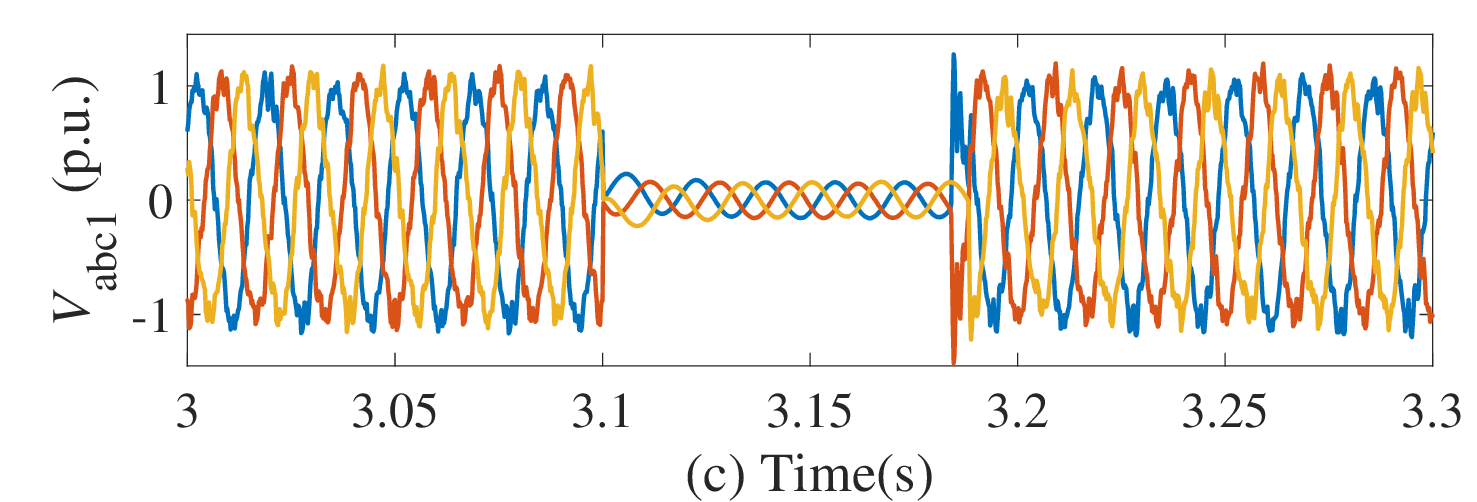}\\
\includegraphics[width=0.485\textwidth]{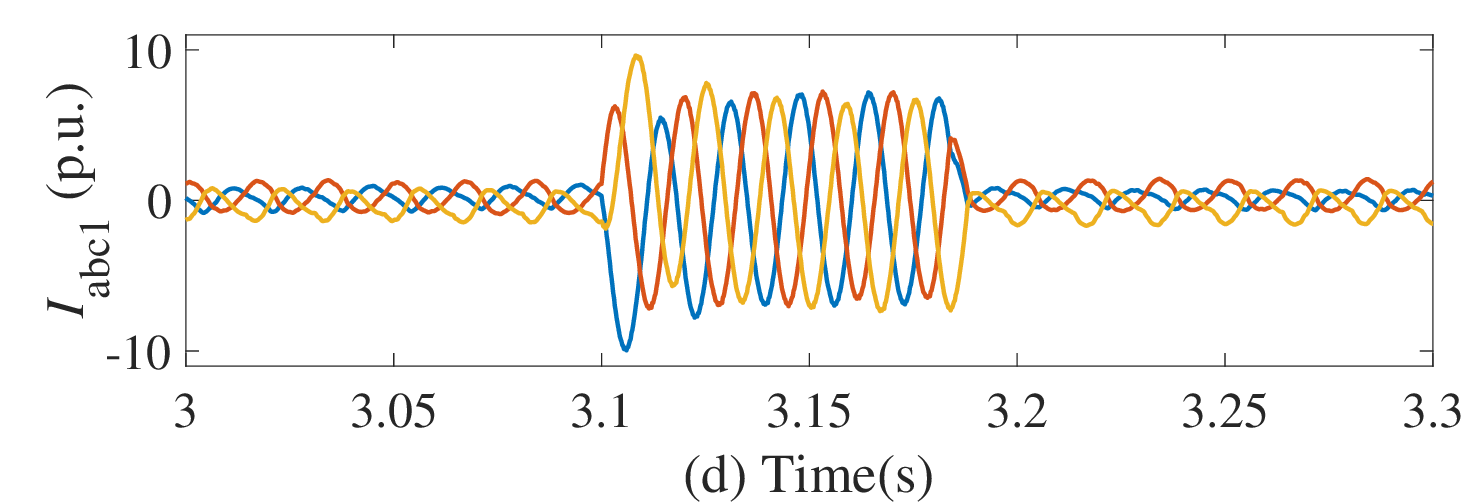}\\
\includegraphics[width=0.485\textwidth]{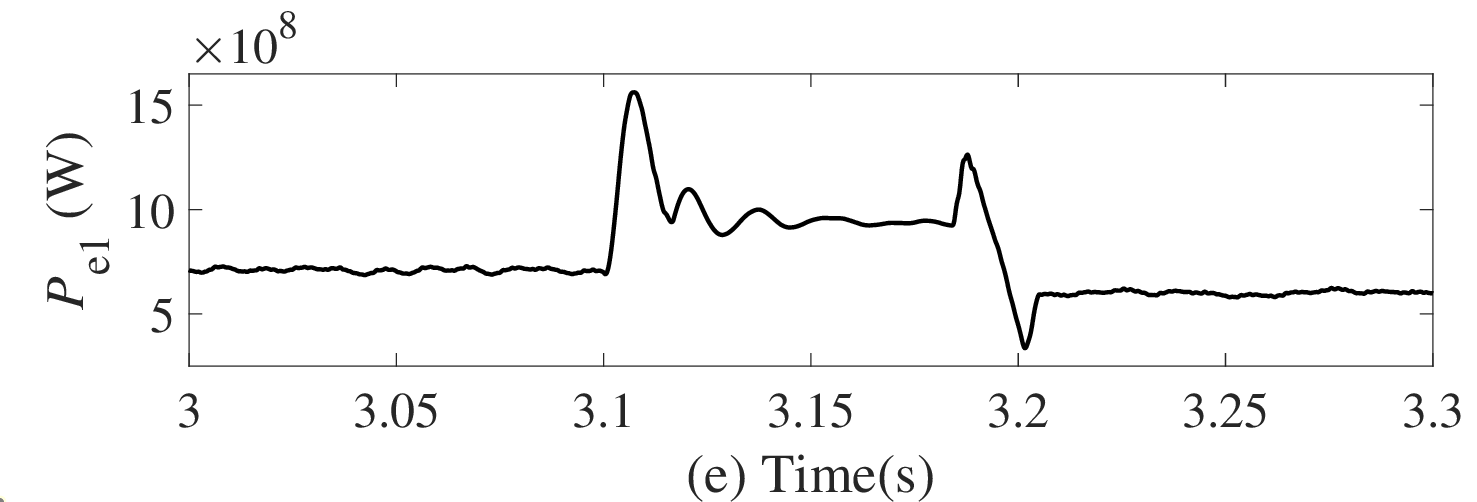}\\
\includegraphics[width=0.485\textwidth]{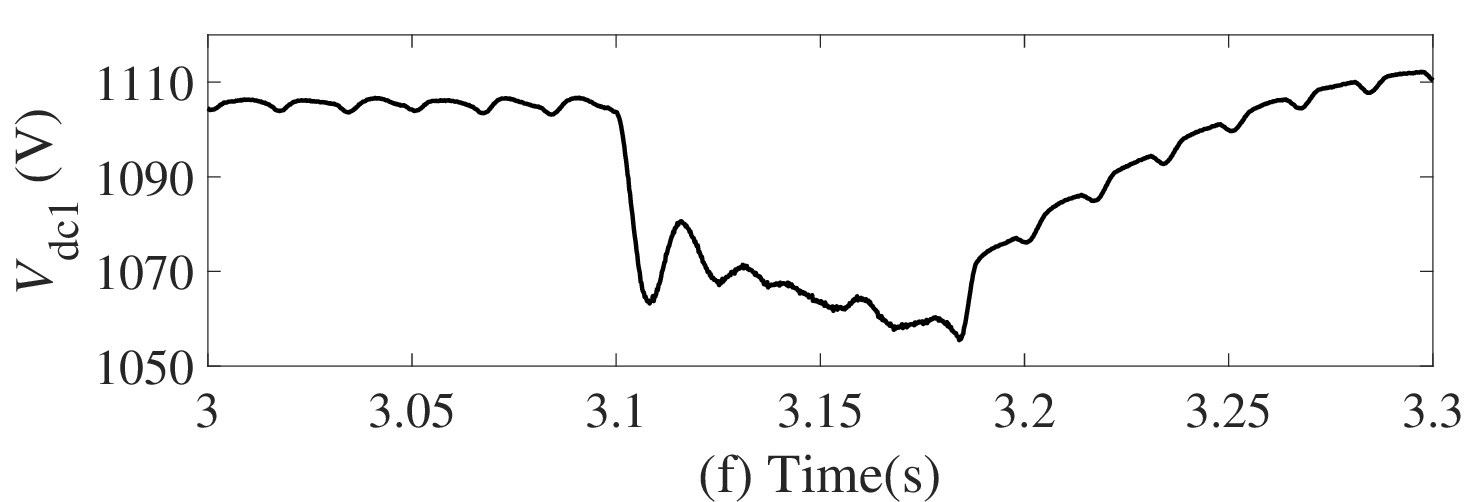}\\
\includegraphics[width=0.485\textwidth]{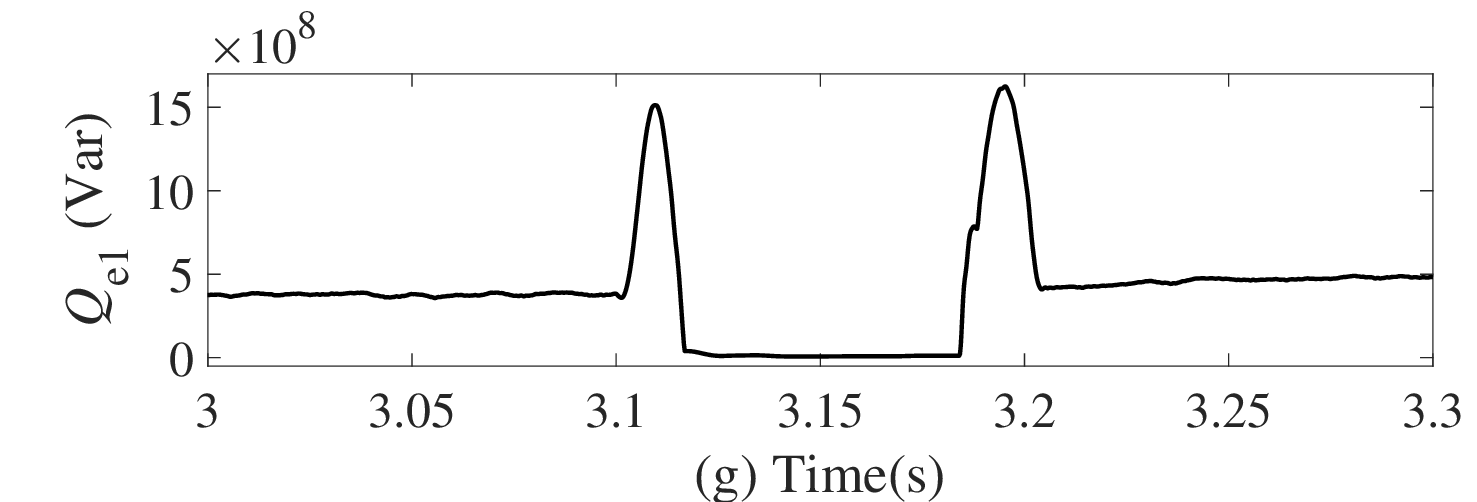}
\caption{Dynamics of WPG$_{1}$ obtained in the case where WPG$_{1}$ operated under no-load condition on the interval $t\in[0,0.5]$s and a three-phase-to-ground fault occurred on bus 5 at $t=3.1$s under the control of AVSCM ((a) Three-phase flux-Linkages measured on generator bus 1 (b) Three-phase currents measured on generator bus 1 on the interval $t\in[0,0.5]$s (c) Three-phase voltages measured on generator bus 1 (d) Three-phase currents measured on generator bus 1 on the interval $t\in[3.1,3.3]$s (e) Active power output of WPG$_{1}$ (f) Capacitor voltage of WPG$_{1}$ (g) Reactive power output of WPG$_{1}$).}
\label{fig_fault2}
\vspace{-0.2cm}
\end{figure}

\begin{figure}[!h]
\centering
\includegraphics[width=0.482\textwidth]{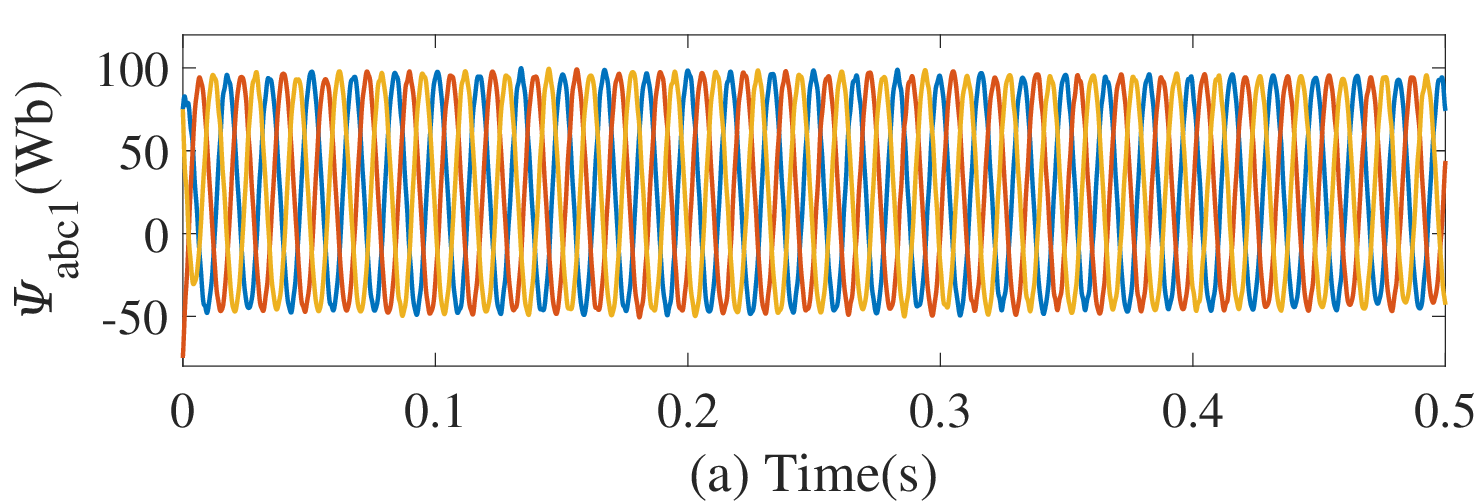}\\
\includegraphics[width=0.482\textwidth]{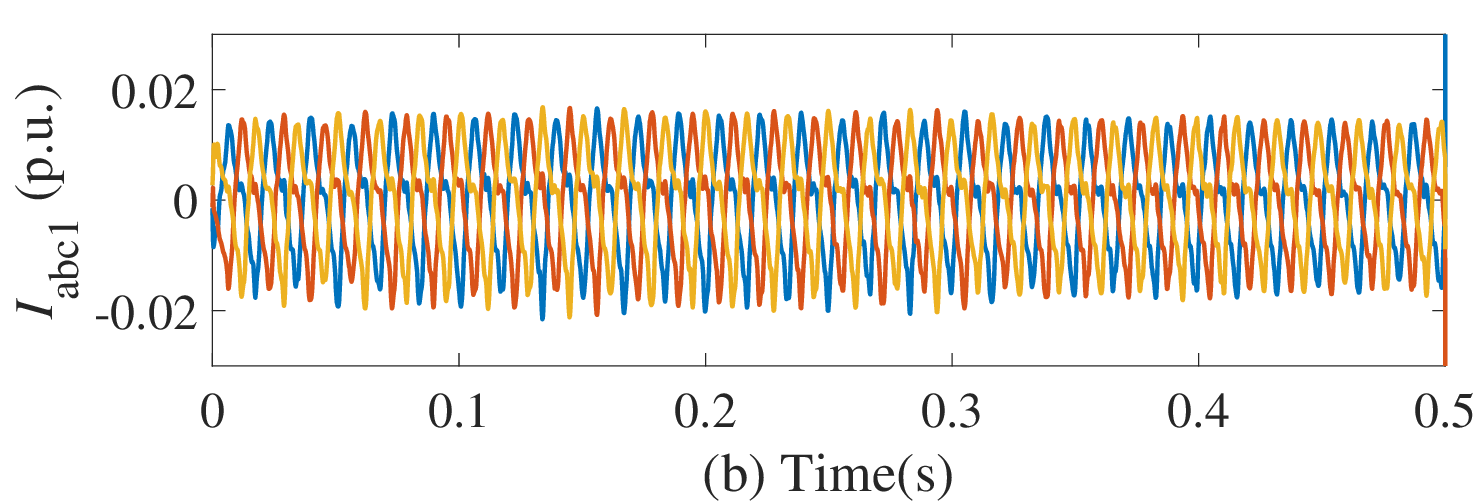}\\
\includegraphics[width=0.482\textwidth]{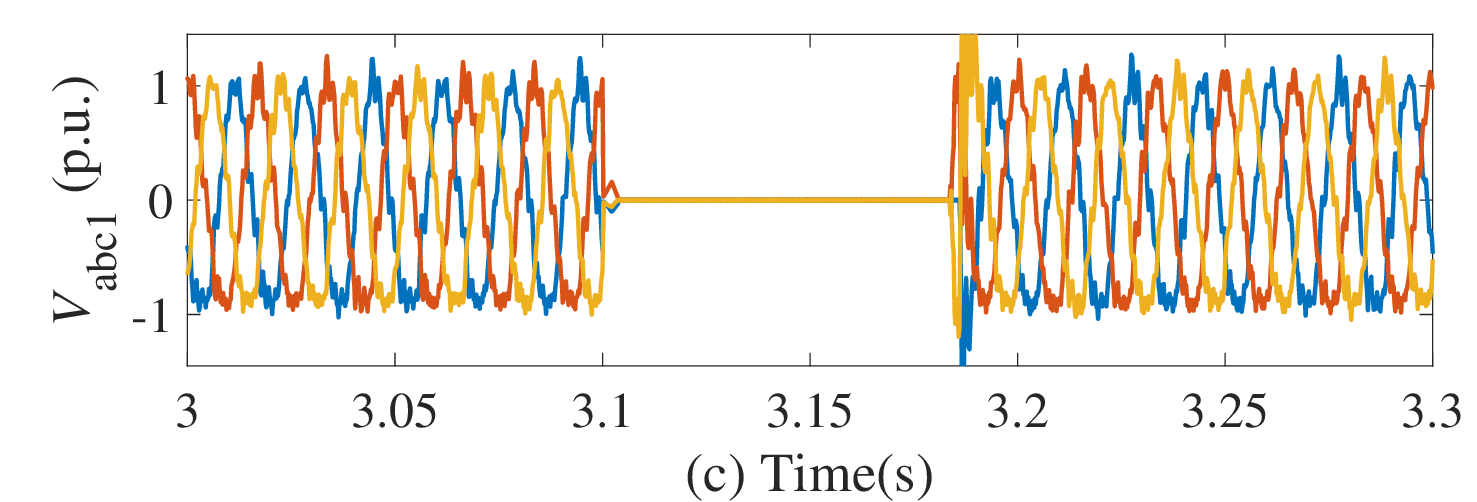}\\
\includegraphics[width=0.482\textwidth]{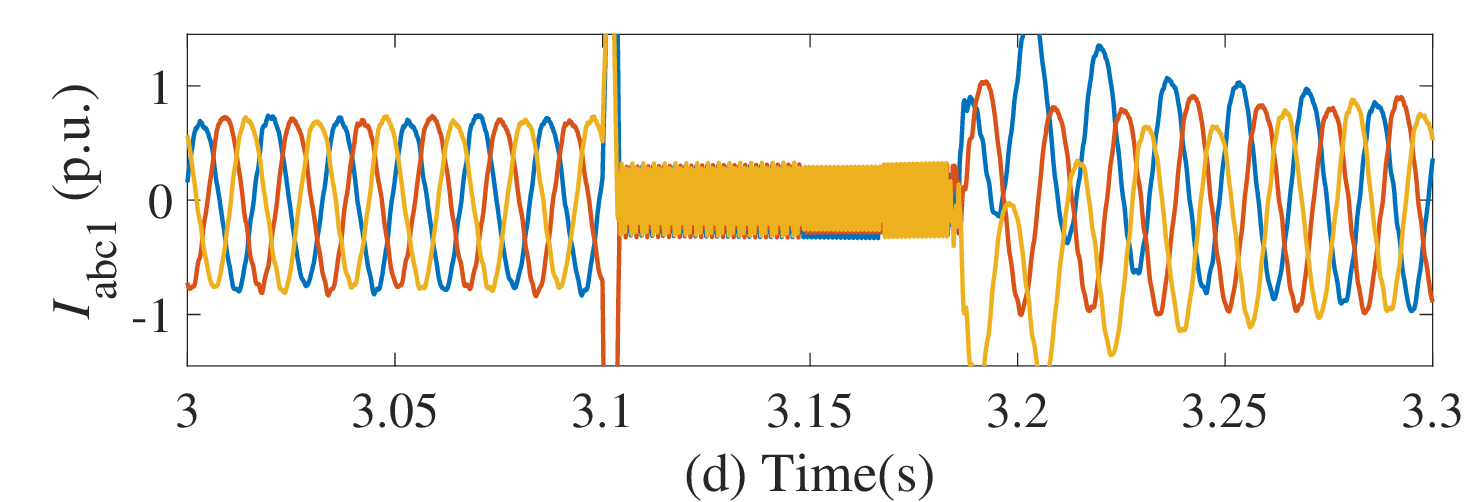}\\
\includegraphics[width=0.482\textwidth]{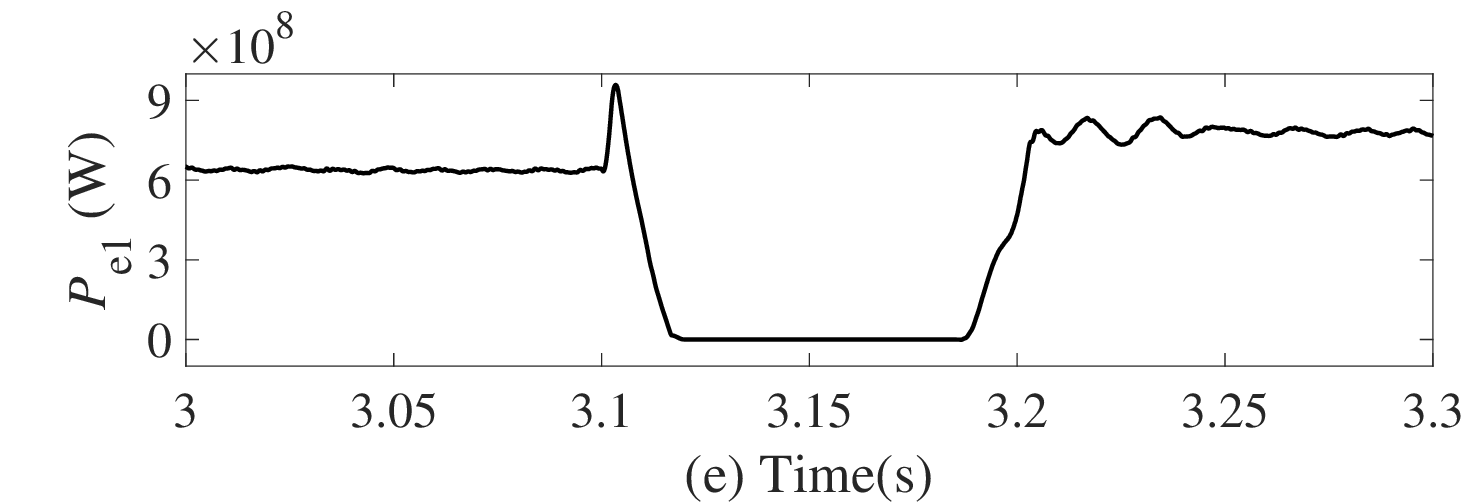}\\
\includegraphics[width=0.482\textwidth]{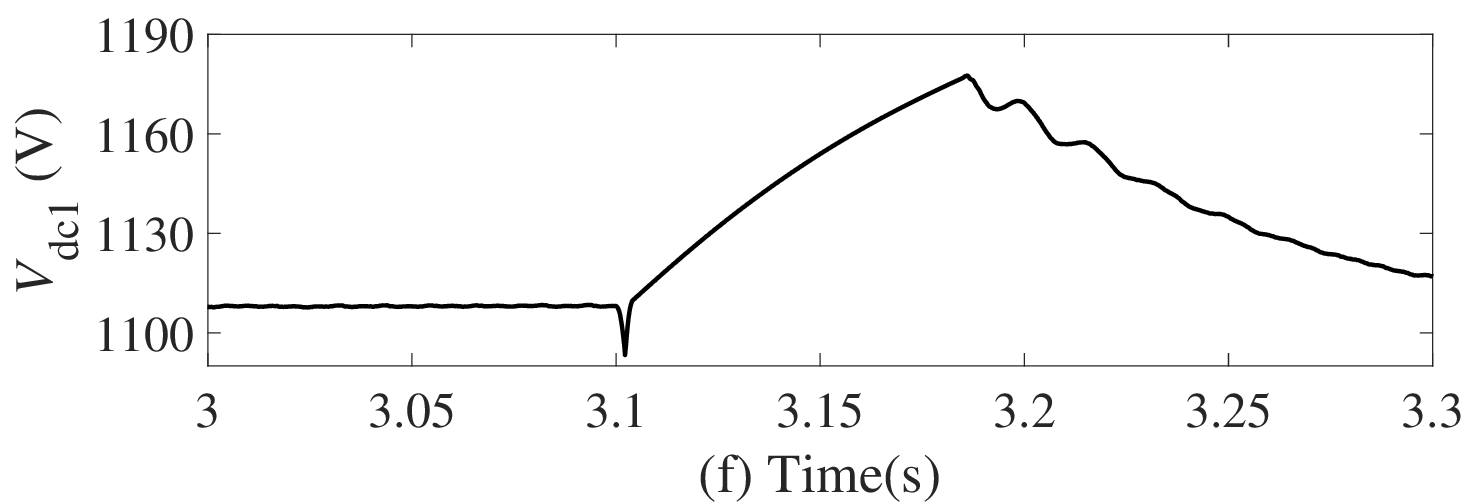}\\
\includegraphics[width=0.482\textwidth]{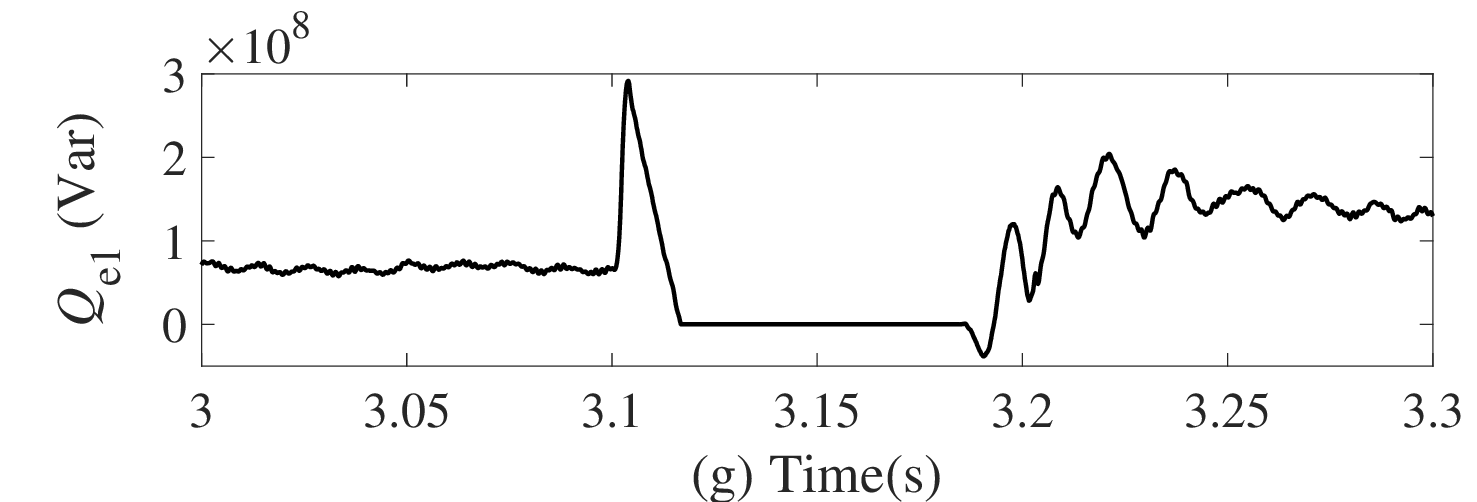}
\caption{Dynamics of WPG$_{1}$ obtained in the case where WPG$_{1}$ operated under no-load condition on the interval $t\in[0,0.5]$s and a three-phase-to-ground fault occurred on bus 5 at $t=3.1$s under the control of NFSCM and LBFC ((a) Flux-linkages measured on generator bus 1 (b) Three-phase currents measured on generator bus 1 on the interval $t\in[0,0.5]$s (c) Three-phase voltages measured on generator bus 1 (d) Three-phase currents measured on generator bus 1 on the interval $t\in[3.1,3.3]$s (e) Active power output of WPG$_{1}$ (f) Capacitor voltage of WPG$_{1}$ (g) Reactive power output of WPG$_{1}$).}
\label{fig_fault1}
\end{figure}

The dynamics of the test system under the control of AVSCM and under the switching control of NFSCM and LBFC, were simulated in the case where WPG$_{1}$ operated under no-load condition on the interval $t\in[0,0.5]$s, and a three-phase-to-ground fault occurred on bus 5. The step-up transformer of WPG$_{1}$ was assumed to have an initial flux-linkage of $[1, -1, 1]$p.u. Since the magnetic flux in the core of the transformer cannot vary in a stepped manner, the mismatch between the flux-linkages of the transformer and the inverter currents brought about irregular components in flux-linkages. These components caused the flux-linkages to be asymmetrical and the core to become magnetically saturated, thereby triggering inrush currents that were over twice the nominal current.

In the system controlled by AVSCM, asymmetrical three-phase flux-linkages were observed on the step-up transformer of WPG$_{1}$ as presented in Fig. \ref{fig_fault2}(a). Inrush currents with a peak value of over 2p.u. were found as shown in Fig. \ref{fig_fault2}(b). By contrast, in the system controlled by NFSCM, the flux-linkages of the step-up transformer of WPG$_{1}$ were brought back to three-phase symmetry rapidly owing to the flux-linkage modulation of NFSCM, as illustrated in Fig. \ref{fig_fault1}(a). As a result, the transformer core did not reach magnetic saturation, and the inrush current of the transformer was avoided, as demonstrated in Fig. \ref{fig_fault1}(b). At $t=0.5$s, the step-up transformer was connected to the external power grid.

\begin{figure}[!t]
\centering
\includegraphics[width=0.485\textwidth]{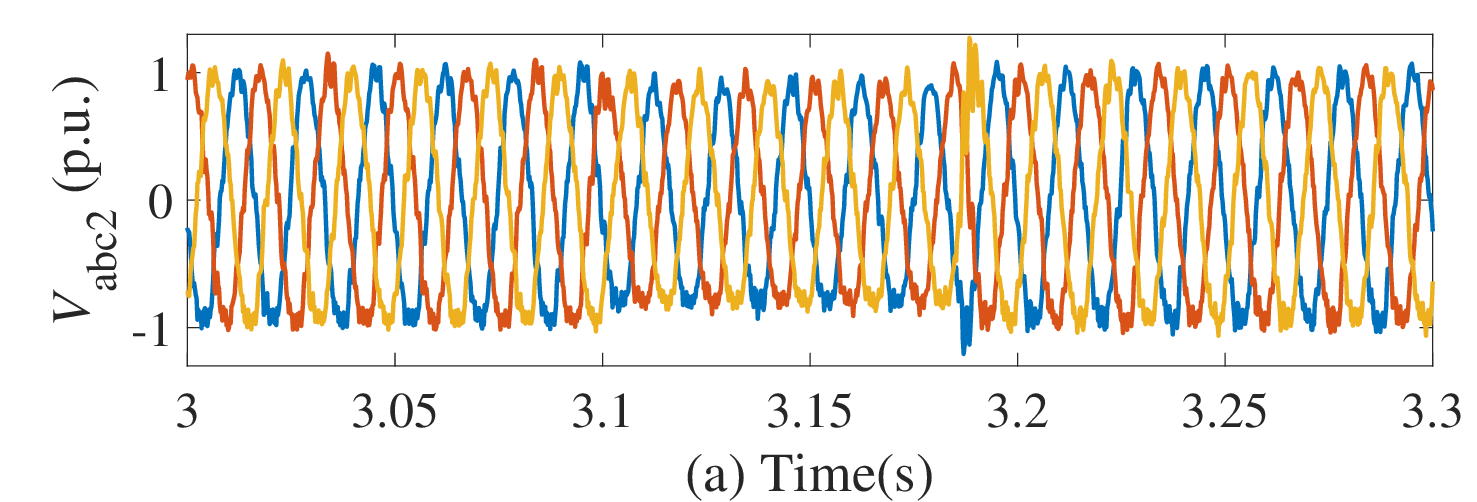}\\
\includegraphics[width=0.485\textwidth]{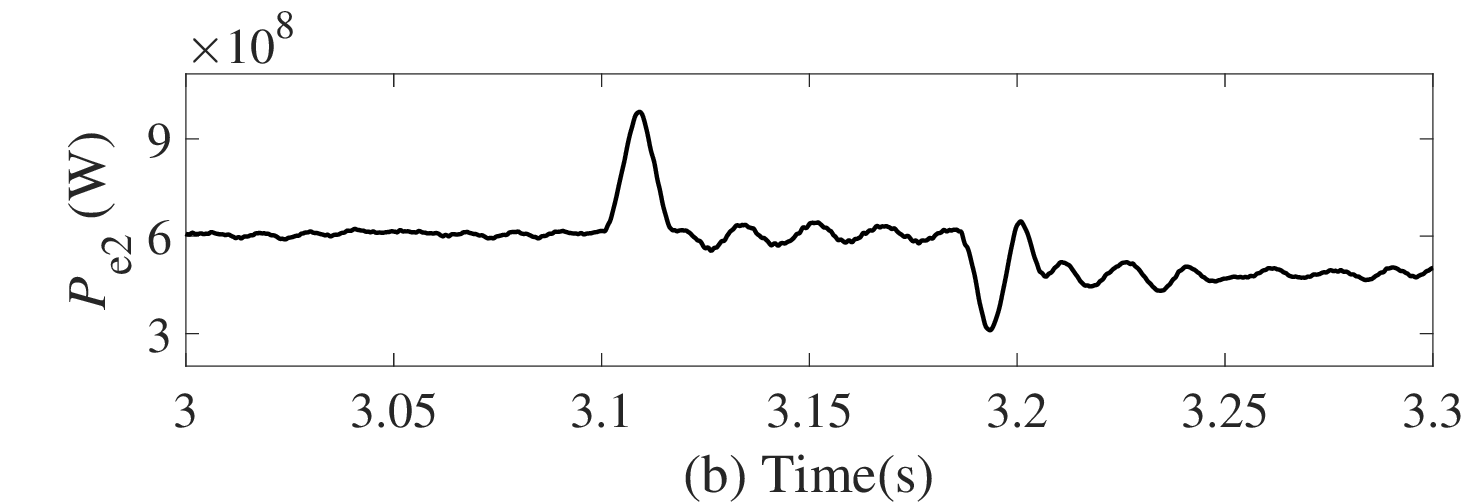}\\
\includegraphics[width=0.485\textwidth]{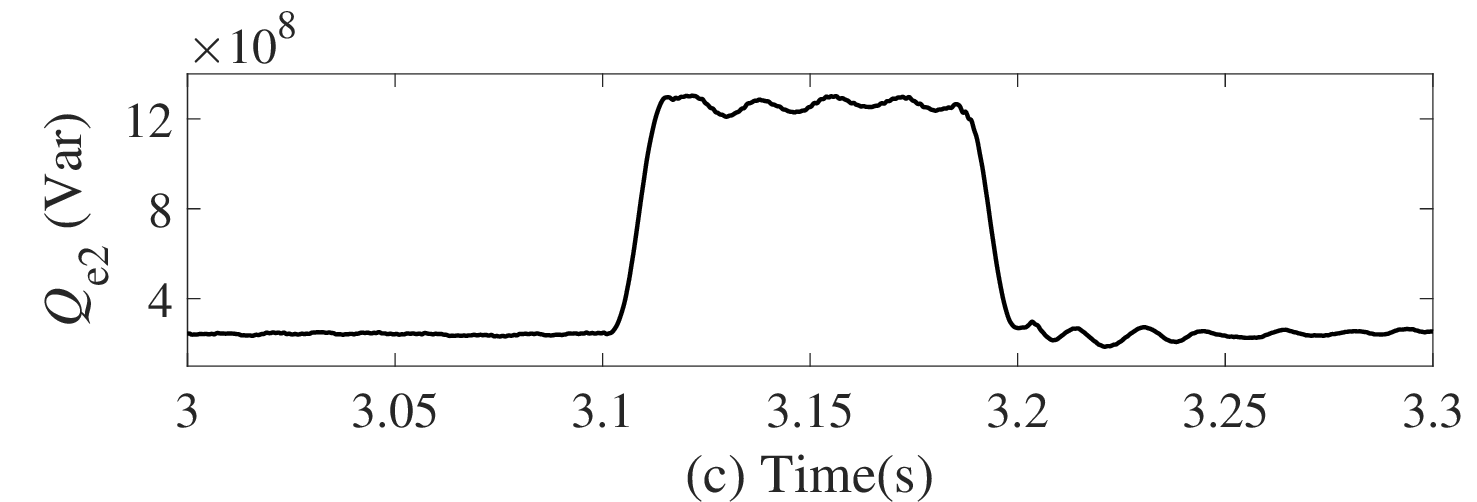}\\
\includegraphics[width=0.485\textwidth]{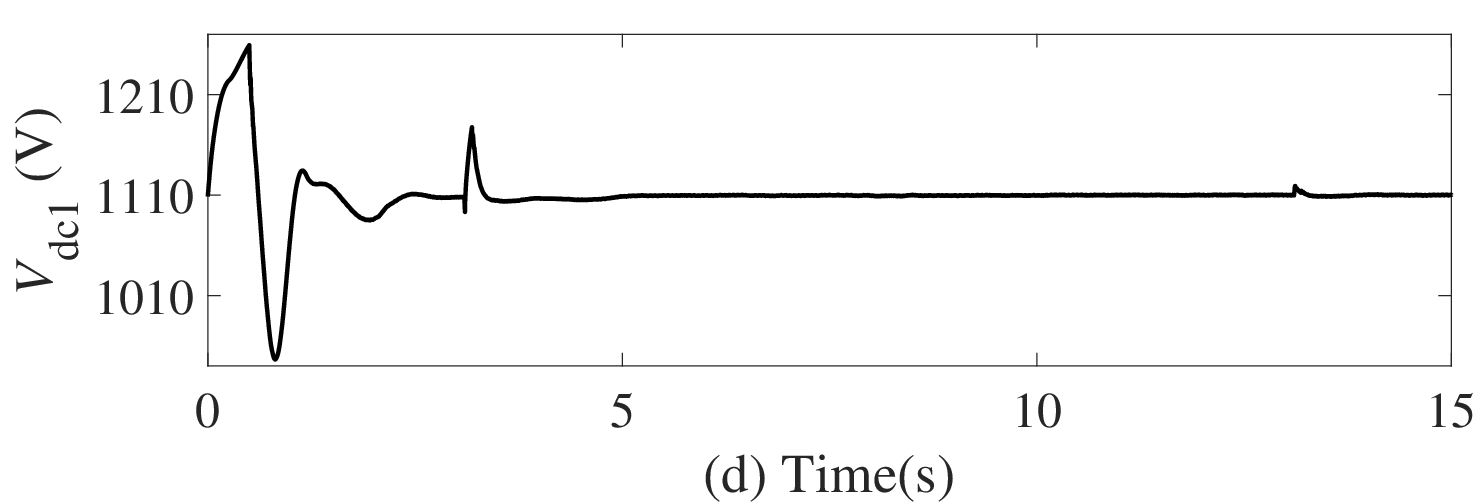}\\
\includegraphics[width=0.485\textwidth]{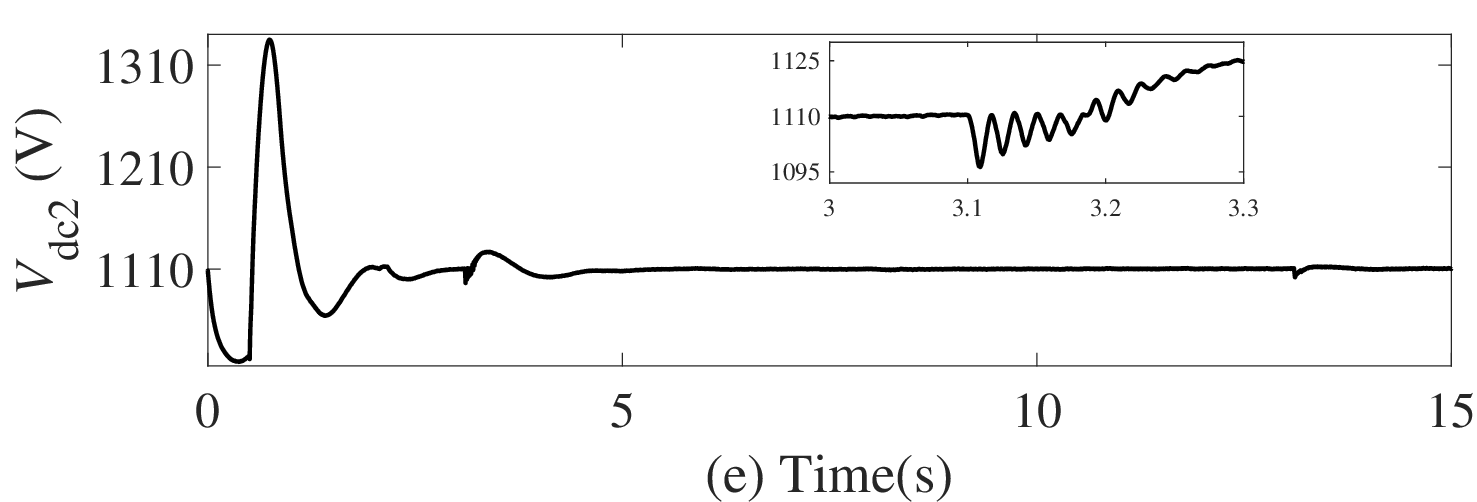}\\
\includegraphics[width=0.485\textwidth]{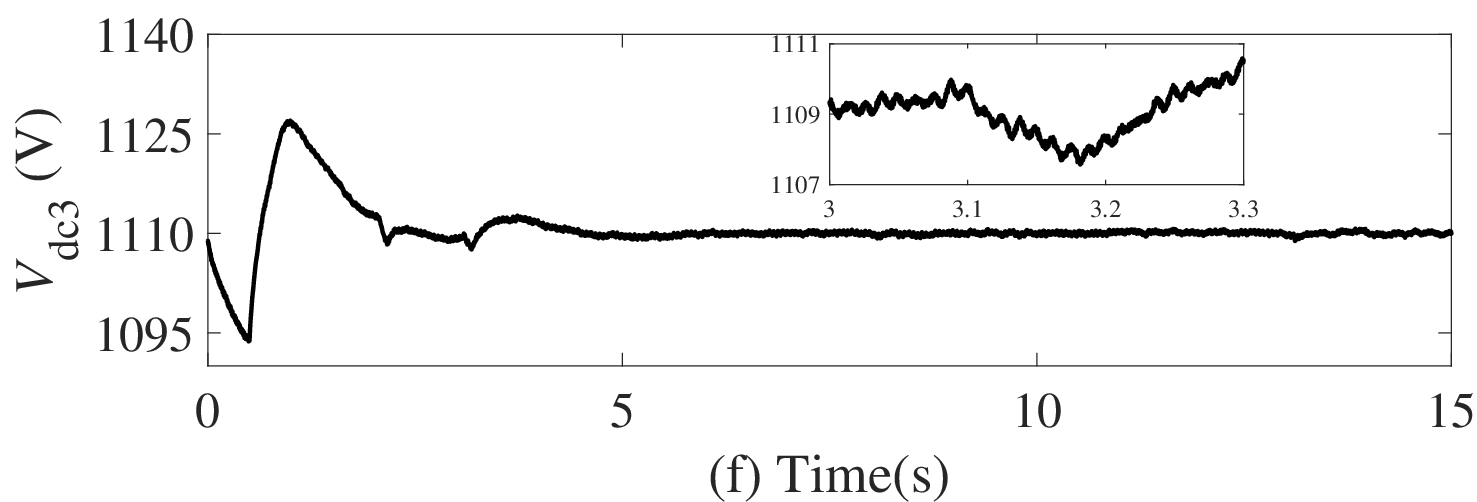}\\
\includegraphics[width=0.485\textwidth]{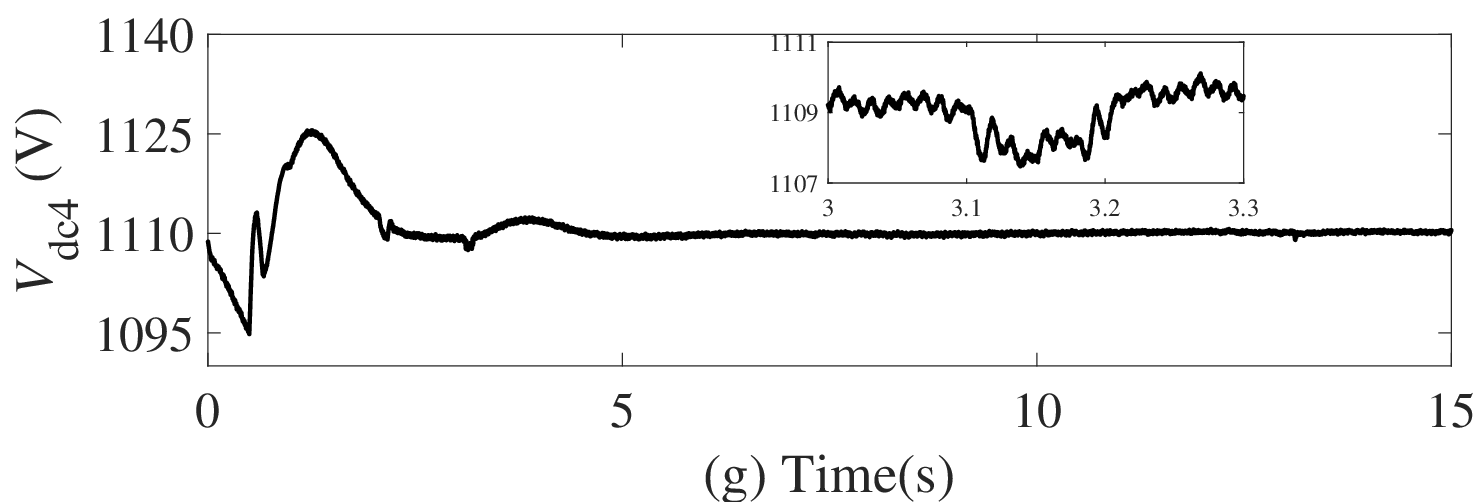}
\caption{Dynamics of WPG$_{2}$, WPG$_{3}$ and WPG$_{4}$ obtained in the case where WPG$_{1}$ operated under no-load condition on the interval $t\in[0,0.5]$s and a three-phase-to-ground fault occurred on bus 5 at $t=3.1$s under the switching control of NFSCM and LBFC ((a) Three-phase voltages measured on generator bus 2 (b) Active power output of WPG$_{2}$ (c) Reactive power output of WPG$_{2}$ (d) Capacitor voltage of WPG$_{1}$ (e) Capacitor voltage of WPG$_{2}$ (f) Capacitor voltage of WPG$_{3}$ (g) Capacitor voltage of WPG$_{4}$).}
\label{fig_fault3}
\vspace{-0.2cm}
\end{figure}

Bus 5 experienced a three-phase-to-ground fault at $t=3.1$s, which was cleared after 5 operation cycles, i.e. 83.3ms. Owing to the fault, the three-phase voltages on generator bus 1 decreased to a low value in the system controlled by AVSCM, as depicted in Fig. \ref{fig_fault2}(c). The short-circuit currents flowing through bus 1 reached 10 times the nominal current, as shown in Fig. \ref{fig_fault2}(d). In practice, such a large current will definitely trigger the over-current protection, and the WPG will be tripped from the grid. The active power output of WPG$_{1}$ increased due to the surge of currents, which further led to the drop in capacitor voltage of WPG$_{1}$, as presented in Fig. \ref{fig_fault2}(e) and Fig. \ref{fig_fault2}(f), respectively. Reactive power output of WPG$_{1}$ declined with the decreasing voltage magnitudes of generator bus 1, as illustrated in Fig. \ref{fig_fault2}(g).

In comparison, the test system controlled by NFSCM and LBFC had much better transient performance. When the short-circuit current was up to the threshold value $\gamma_{i}$, the inverter control was switched from NFSCM to LBFC, which regulated the current into the error funnels $[\varphi^{-}_{ji}, \varphi^{+}_{ji}]$, as presented in Fig. \ref{fig_fault1}(d). Meanwhile, the voltage of bus 1 was controlled at a lower level along with the operation of LBFC, as shown in Fig. \ref{fig_fault1}(c). Therefore, the fault current suppression prevented the inverter from being tripped by the over-current protection. The active and reactive power outputs of WPG$_{1}$ dropped to zero during the fault, which is illustrated in Fig. \ref{fig_fault1}(e) and Fig. \ref{fig_fault1}(g), respectively. It followed that the capacitor voltage of WPG$_{1}$ increased, as presented in Fig. \ref{fig_fault1}(f).

According to Fig. \ref{fig_fault3}(a), voltages of the PCCB of WPG$_{2}$, which is farther from the fault point than WPG$_{1}$, presented less magnitude drop. In order to sustain the PCCB voltages, the exciter-mimicking loop of NFSCM enhanced the magnitude of the flux-linkage and there was a boost in the reactive power output of WPG$_{2}$, as detailed in Fig. \ref{fig_fault3}(c). WPG$_{2}$ still had a mild fluctuating increase in active power output due to inertial response, which can be observed in Fig. \ref{fig_fault3}(b). WPG$_{3}$ and WPG$_{4}$ presented similar dynamics to WPG$_{2}$. Fig. \ref{fig_fault3}(e), \ref{fig_fault3}(f), and \ref{fig_fault3}(g) display the dynamics of the capacitor voltages of WPG$_{2}$, WPG$_{3}$, and WPG$_{4}$, respectively. These capacitor voltages dropped to various nadirs when the fault happened, as illustrated in Fig. \ref{fig_fault3}(d)\,-\,\ref{fig_fault3}(g). Nevertheless, once the perturbation subsided, all of them returned to a common value and reached a novel synchronized state.

\section{Conclusions}\label{sec_conclusion}

This paper has proposed a CVBS-based NFSCM system for the power system with 100\% wind power generation. Inverters of WPGs are controlled to be flux-linkage sources, and the angular position of the flux-linkage source moves according to the equations of motion of capacitor voltages. With CVBS-based NFSCM, the 100\% wind power generation system achieves self-synchronization, inertia response, and primary frequency regulation. The control system composed of NFSCM and LBFC not only resolves the inrush current issue, but also enhances the fault current rejection capability of WPGs.

Simulation results, obtained in the load increase case, have verified the flux-linkage synchronizing performance of WPGs. Energy storages of WPGs were able to provide primary frequency regulation responding to the load increase of the external power grid, which further enabled the capacitor and SG of the WPG to offer inertial support to the drop of capacitor voltages. NFSCM eliminated the DC components of inverter currents that appeared under the control of AVSCM. This validates the superiorities of operating the inverter as a flux-linkage source over operating it as a voltage source.

Referring to the simulation results, obtained in the case of no-load operation of WPG$_{1}$ and grid fault, NFSCM's precise control of the flux-linkage avoided the intimate magnetic saturation of the transformer, thereby removing the excitation inrush currents. The switching from NFSCM to LBFC after the inverter current reached the threshold value effectively restrained the fault currents, and prevented the inverter from being tripped from the grid. Moreover, it was validated that the synchronization of WPGs under the control of CVBS-based NFSCM was indicated by the capacitor voltage of DC-links, which meets the analytical conclusions reached by (\ref{equ_voltage_synchronization}). Therefore, CVBS-based NFSCM provides a novel synchronizing mechanism for 100\% power electronics-interfaced power systems. The secondary frequency regulation and the optimal configuration and control of energy storages in power systems with 100\% renewable generation will be a topic of future research.

\bibliographystyle{IEEEtran}
\bibliography{Ref}
\appendices

\end{document}